\RequirePackage{fix-cm}
\documentclass[smallextended]{svjour3}       
\smartqed  
\usepackage[hidelinks]{hyperref}
\usepackage{color}
\usepackage{float}
\usepackage{braket}
\usepackage{amsmath}
\usepackage{graphicx}
\usepackage{caption}
\usepackage{subcaption}

\usepackage[misc,geometry]{ifsym}

\usepackage{algorithm}
\usepackage{algorithmic}
\usepackage{mathptmx}      
\journalname{Quantum Information Processing}
\begin{document}

\title{Design of a Universal Logic Block for Fault-Tolerant Realization of any Logic Operation in Trapped-Ion Quantum Circuits}
\titlerunning{Design of a ULB for Fault-Tolerant Realization of any Logic Operation in Trapped-Ion Quantum Circuits}


\author{H. Goudarzi \and
		M. J. Dousti \and \\
		A. Shafaei Bejestan \and
		M. Pedram}

\authorrunning{H. Goudarzi et al.} 

\institute{H. Goudarzi (\Letter) \and M. J. Dousti \and A. Shafaei \and M. Pedram \at
              Department of Electrical Engineering, University of Southern California,  Los Angeles, CA, USA  \\
              \email{hgoudarz@usc.edu}\\
              \\
              M. J. Dousti\\
              \email{dousti@usc.edu}\\
              \\
              A. Shafaei\\
              \email{shafaeib@usc.edu}\\
			  \\
              M. Pedram\\
              \email{pedram@usc.edu}\\
}


\maketitle

\begin{abstract}
This paper presents a physical mapping tool for quantum circuits, which generates the optimal Universal Logic Block (ULB) that can, on average, perform any logical fault-tolerant (FT) quantum operations with the minimum latency. The operation scheduling, placement, and qubit routing problems tackled by the quantum physical mapper are highly dependent on one another. More precisely, the scheduling solution affects the quality of the achievable placement solution due to resource pressures that may be created as a result of operation scheduling  whereas the operation placement and qubit routing solutions influence the scheduling solution due to resulting distances between predecessor and current operations, which in turn determines  routing latencies. The proposed flow for the quantum physical mapper captures these dependencies by applying (i) a loose scheduling step, which transforms an initial quantum data flow graph into one that explicitly captures the no-cloning theorem of the quantum computing and then performs instruction scheduling based on a modified force-directed scheduling approach to minimize the resource contention and quantum circuit latency, (ii) a placement step, which uses timing-driven instruction placement to minimize the approximate routing latencies while making iterative calls to the aforesaid force-directed scheduler to correct  scheduling levels of quantum operations as needed, and (iii) a routing step that finds  dynamic values of routing latencies for the qubits. In addition to the quantum physical mapper, an approach is presented to determine the single best ULB size for a target quantum circuit by examining the latency of different FT quantum operations mapped onto different ULB sizes and using information about the occurrence frequency of operations on critical paths of the target quantum algorithm to weigh these latencies. Experimental results show an average latency reduction of about 40\% compared to previous work.


\end{abstract}
\clearpage
\section{Introduction}
\label{intro}

\noindent It has been speculated that quantum computers, which harness physical phenomenon unique to quantum mechanics (especially quantum interference and superposition) to realize a fundamentally new mode of information processing, can overcome the limitation of classical computers in efficiently solving hard computational problems. In particular, polynomial-time algorithms have been offered to solve several NP-intermediate problems by using quantum computers \cite{nielsen_quantum_2011}.
 
A key challenge in building large-scale quantum computers is the environmental noise - more precisely, loss of quantum information due to unwanted interactions between the quantum system and its environment. This phenomenon is called \textit{quantum} \textit{decoherence}. It has been shown that arbitrarily accurate quantum computation is possible, provided that the error per operation is below a \textit{threshold} value \cite{preskill_reliable_1998} and \cite{knill_resilient_1998}. In other words, quantum error correction works if the level of imperfections and noise in quantum gate operations and measurements are below a certain \textit{error threshold} and if corrections can be applied repeatedly \cite{schindler_experimental_2011}. Based on the error threshold for each quantum circuit fabric technology, an appropriate encoding technique (such as m-level concatenated Steane code \cite{steane_multiple-particle_1996}) is selected to achieve fault tolerant realization of the quantum circuit. 

A typical quantum circuit fabric consists of a two-dimensional array of identical primitive structures (called \textit{primitive cells} in this paper), each structure containing some sites for generating/initializing qubits, measuring them, performing operations on one or two qubits, and channels for moving qubits or swapping their information. See fig. \ref{fig:fig3} for an example of such an array of primitive cells. Unfortunately, dealing directly with this primitive cell array is very cumbersome and unwieldy. So in practice another 2-D array of super-templates (which we call \textit{tiles}) is built. Each tile comprises a number of primitive cells. Instead of mapping a quantum circuit to the quantum fabric, the quantum circuit is mapped to this tiled architecture (see below).

A quantum logic synthesis tool (surveyed in reference \cite{saeedi_synthesis_2011}) generates a quantum circuit. Every qubit in the output is called a \textit{logical qubit}, which is subsequently encoded into several \textit{physical qubits} to detect and correct potential errors. To prevent the propagation of errors in the quantum circuit, the (reversible) \textit{logic gates} in the synthesized circuit (which are typically, NOT, CNOT and Toffoli gates) must be converted into FT\textit{ quantum operations}. A possible universal (but redundant) set of FT quantum operations includes CNOT, H (Hadamard), T ($\pi $/4 rotation), S (phase), X, Y and Z gates. Implementation of these FT quantum operations depend on the chosen error correction method. Note that the set \{CNOT, H, T\} constitutes a universal basis for quantum circuit realization--the other instructions are included to allow logic simplification during the process of converting the logic synthesis output to the FT quantum operation realization. A given quantum circuit fabric is natively capable of performing a universal set of one and two-qubit instructions (also called \textit{native quantum instructions}). This set differs among various quantum technologies (e.g., ion-trap vs. superconducting qubit vs. quantum dots). Each FT quantum operation can be implemented by using a composition of these native quantum instructions. The transformation from logical gates (which are the result of the quantum logic synthesis) to the FT quantum operations and from the FT quantum operations to the native quantum instructions are called \textit{FT quantum synthesis} and \textit{native quantum synthesis}, respectively. FT and native quantum syntheses are outside of the scope of the present paper.

Each of the FT quantum operations performs a desired function on one or two logical qubits as the input producing one or two logical qubits as the output; each of the input qubits is encoded with some number of physical qubits. The output qubits will also be encoded. Moreover, each of these FT quantum operations requires subsequent \textit{syndrome extraction} circuitry to detect and correct any errors (up to a certain number) that may have been introduced by the quantum operation itself.

\begin{figure}[h]
\centering
\scalebox{1}{\includegraphics*[width=0.5\textwidth]{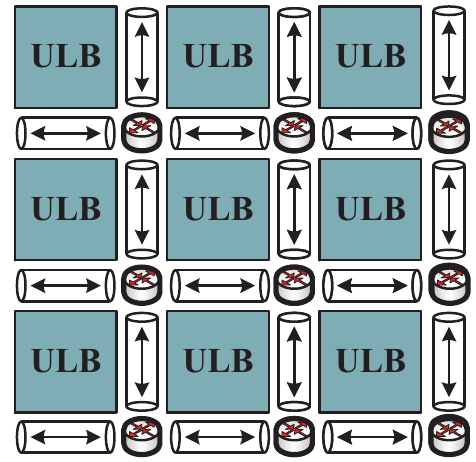}}
\caption{Tile-based architecture}
\label{fig:fig1}
\end{figure}

Based on the adopted encoding scheme, implementation of each of the FT quantum operations may require hundreds to tens of thousands of native quantum instructions. Different works in the literature e.g., \cite{metodi_quantum_2005} and \cite{whitney_fault_2009} have suggested using the \textit{tiled quantum architecture} (TQA), comprised of a regular two-dimensional array of \textit{Universal Logic Blocks} (ULBs) to avoid dealing with this complexity. Notice that each ULB in the TQA is capable of performing \textit{any} FT quantum operation. ULBs are separated by routing channels, which are used to move logical qubits (or information about these qubits) from some source ULBs to a target ULB in the TQA. A pictorial representation of the TQA is shown in fig. \ref{fig:fig1}. The quantum structures placed at the junctions of routing channels may be thought of as \textit{quantum crossbars} (possibly with some \textit{qubit purification} \cite{cirac_optimal_1999} capability). Routing channels and quantum crossbars are also built from primitive cells.

A ULB is analogous to a \textit{Configurable Logic Block} (CLB) in a Field Programmable Logic Array (FPGA) device, in that it can implement any of a set of desired functions. Moreover, the same ULB (as identified by its unique row and column indices in the ULB array) can be configured to perform different FT quantum operations at different times as needed. This is analogous to an \textit{on-the-fly-reconfigurable CLB}.  

Trapped-ion technology is a very promising technology for implementing quantum circuits \cite{ladd_quantum_2010}. Hence, it is selected as the underlying technology for our quantum physical mapping tool. See section \ref{qcrep} for more details of this technology. A trapped-ion quantum circuit fabric comprises of small quantum cells that are organized in a two-dimensional grid, similar to the TQA. Although each primitive cell has the complete set of capabilities provided by the trapped-ion technology, it is not large enough to realize a complete FT operation with an acceptable latency. That is why we resort to the ULB abstraction. Notice that each ULB itself is simply a fixed-size 2-D array of primitive cells, which can realize any FT operation with an acceptable latency.  

After appropriate high-level transformations, a quantum algorithm may be represented by a \textit{quantum operation dependency graph} (QODG), where nodes represent FT quantum operations and edges capture data dependencies. Based on the target quantum fabric and error threshold, a particular quantum coding scheme is selected, and subsequently, a high-level tool maps the QODG into a TQA, where each ULB (tile) in this architecture can implement any desired one or two-input quantum operation in a fault-tolerant way. 
The latency of the quantum algorithm mapped to the TQA can be calculated as the length of the longest path in the QODG, where the length of an input-output path in the QODG is the summation of operation latencies plus routing latencies. The operation latencies are dependent on how precisely the required operation is realized within a ULB. This realization comprises static provisioning (setting the size of the ULB in terms of an array of primitive structures in the target quantum fabric) and mapping of the target operation into the ULB. On the other hand, the routing latencies are dependent on how long it takes for the logical qubits to move from source ULBs to a target ULB to perform the FT quantum operation. The routing latency, which depends on the actual move distances and the latency of moving a qubit through some link in the TQA, is highly sensitive to ULB size, operation placement, and qubit routing on the TQA. 

The focus of this paper is on finding the optimal ULB size for the TQA in the trapped-ion technology. The most important part of ULB design is to minimize the operation latency for a fixed ULB size by optimally mapping the quantum instructions to the underlying quantum fabric. This action is called \textit{quantum physical mapping}. The quantum physical mapping procedure is decomposed into three sub-procedures: (a) quantum instruction scheduling, (b) quantum instruction placement, and (c) dynamic qubit routing. Next, based on operation usage frequencies in the target quantum algorithm and the average movement of logical qubits  per operation in the TQA, the optimal size of the ULB is determined. 

Each quantum fault-tolerant operation and its required syndrome extraction circuit can be modeled by a \textit{quantum instruction dependency graph} (or QIDG). A QIDG for an FT quantum operation is what a QODG is to a quantum algorithm. As an example, a CNOT operation between two physical qubits may be implemented by using more than one native quantum instructions in the trapped-ion technology. Moreover, this number of native quantum instructions is multiplied by a factor of seven if 7-qubit Steane code [[7,1,3]] is used to implement the CNOT in a fault-tolerant manner. Adding syndrome extraction circuit for logical qubits increases the number of native quantum instructions even more. These numbers are far larger in the case of more complex (i.e., non-transversal) FT quantum operations such as T operation in the [[7,1,3]] Steane code, or more complex quantum encoding schemes such as the \textit{concatenated} 7-qubit Steane code.

Different solutions for quantum physical mapping are introduced in the literature -- see for example, \cite{metodi_quantum_2005,whitney_fault_2009,balensiefer_quale:_2005,metodi_scheduling_2006,whitney_automated_2007,maslov_quantum_2008,mohammadzadeh_auxiliary_2011,dousti_minimizing_2012}. In this work, an iterative approach is proposed to schedule and place the instructions on quantum fabric to capture the relation between scheduling level of the instructions, instruction placement solution and qubit routing latencies. The proposed algorithms in each step are based on the state-of-the-art scheduling and placement solutions adding the unique characteristics of the quantum instructions scheduling and placement.
To capture the quantum no-cloning theorem that forbids fan-out in quantum circuits, a pre-processing step is implemented on the QIDG to resolve the multiple read dependencies between instructions by adding auxiliary edges between them considering the criticality factor of each instruction. A force-directed instruction scheduling approach is used as the scheduling algorithm to minimize the contention, i.e. the number of concurrent quantum instructions, and hence reduces the routing latency due to congestion among qubits. Quantum instructions can only be executed at specific locations on the quantum circuit fabric called \textit{interaction wells}. The placement of the instructions in these interaction wells affects their start time due to the qubit routing latencies. Hence, we present a placement approach based on a state-of-the-art placement engine with adaptive re-levelization of the operations. More precisely, the placement algorithm calls the quantum instruction scheduler several times to update the scheduling levels of instructions based on the current quantum placement solution. These algorithms are implemented as part of a quantum CAD tool called \textit{quantum ULB factory designer} (QUFD). 

To evaluate the performance of the proposed quantum instruction scheduling and placement algorithms, a quantum fabric emulator including a greedy router is developed. The result of the proposed algorithms with respect to previous work is evaluated. On average, 40\% improvement with respect to the state-of-the-art quantum mapper is observed. Moreover, considering Toffoli profiling information, the best aspect ratio for a ULB to achieve the best latency is calculated and explained in the experimental section.

The remainder of this paper is organized as follows. Reviews of the most relevant prior work are presented in section \ref{prior}. Background on the quantum circuit representation and trapped-ion fabric is presented in section \ref{qcrep}. The proposed flow for ULB design is presented in section \ref{cadflow}. Quantum physical mapping tool is discussed in section \ref{physicalmapping}. Quantum instruction scheduling and placement problem formulation and proposed algorithms are presented in sections \ref{sched} and \ref{placement}. Experimental results are presented in section \ref{results} and the paper is concluded in section \ref{concl}.

\section{Prior Work}
\label{prior}

QUFD is not the first tool of its kind, but it dramatically improves the performance of existing tools. The rest of this section gives a survey on previous quantum physical design tools and summarizes the contributions and unique features of QUFD compared to the prior art.

Balensiefer et al. \cite{balensiefer_quale:_2005} developed QUALE, a tool for designing microarchitectures for trapped-ion quantum computers. This tool employs as late as possible scheduling on QODG, greedy qubit placement, and Pathfinder \cite{mcmurchie_pathfinder:_1995} based routing. The key disadvantage of the proposed placement in \cite{balensiefer_quale:_2005} is that it is independent of the structure of the given QODG. Hence, two qubits that have a lot of interactions may be placed far from each other. This increases the routing cost of bringing the two qubits together to perform the required gate level operations. Moreover, it increases the congestion in the routing channels.

QPOS \cite{metodi_scheduling_2006} uses a similar flow as QUALE, but distinguishes between the source and destination operands of a two-qubit instruction during the routing step. This tool employs a combination of list scheduling and as soon as possible scheduling methods for quantum instruction scheduling and a priority-based routing algorithm for qubit routing. Reference \cite{whitney_automated_2007} tweaks QPOS by creating a priority queue for instructions based on the total delay of their dependent instructions. Reference \cite{whitney_fault_2009} develops a quantum mapping tool and compares different architectures for quantum fabrics. The main focus of \cite{whitney_fault_2009} is to improve the quality of high-level mapper. However, it lacks the aggressive optimization aspect of tile (ULB) design. We believe that this optimization dramatically affects the total latency of the final quantum circuit.

Reference \cite{maslov_quantum_2008} discusses the notion of the classical and quantum placement including differences and similarities. The authors propose heuristics to place the quantum instructions and route the qubits by applying the SWAP instruction to minimize the total latency of the quantum circuit. The proposed heuristics in \cite{maslov_quantum_2008} consider fixed instruction scheduling solution and does not consider the global instruction placement solution. Fixed instruction scheduling means the order of instructions are determined before determining the placement and routing solutions which limits the effect of scheduling algorithm solution on the circuit's latency. Moreover, considering the global placement solution reduces the possibility of being trapped in a local optimum. Reference \cite{mohammadzadeh_auxiliary_2011} proposes a physical synthesis flow for ancillary qubit selection to reduce the total latency of the quantum physical mapping. This work is focused on changing the design of the fabric to decrease the circuit latency (similar to \cite{whitney_automated_2007}).

A mapping tool for trapped-ion technology, called QSPR, is proposed in \cite{dousti_minimizing_2012}. This tool supports ion multiplexing in the trapped-ion technology; it offers a more optimized global placement of the qubits on the quantum circuit fabric by using an iterative placement approach based on forward and backward computations in a QIDG; and it improves the routing solution by simultaneously moving the source and destination qubits toward a designated site. It provides about 41\% improvement over QUALE. We consider this mapping tool the-state-of-the art technique and compare our results with it. The proposed greedy method for instruction placement in this approach cannot decrease the latency in large quantum circuits because the impact of initial optimized qubit placement fades after running of few instructions; in other words, qubit placement can only enhance the routing delay for some of the first instructions and has no effect on later instructions. 

The proposed quantum physical mapper and ULB designer in our work aim to automate ULB design in quantum mapping tools to decrease the total latency of the quantum circuit realization as much as possible. The proposed quantum physical mapper considers the effect of the routing time on quantum instruction scheduling and placement and presents a cross-layer optimization between these solutions to minimize the latency. In addition, the proposed offline method to optimize the ULB size considering the logical qubit routing times and operation repetition frequency on the critical path of the quantum circuit is not discussed in the prior art.

\section{Quantum circuit representation and trapped-ion technology}
\label{qcrep}

\subsection{ Quantum circuit and its representation}

Quantum circuits can be composed in Quantum Assembly (QASM) language \cite{balensiefer_evaluation_2005}, which provides a compact way of representing quantum circuits. QASM is a low-level quantum language that does not support complex control and data structures like arrays, loops, conditionals, etc. 

\begin{figure}
        \centering
        \begin{subfigure}[b]{0.4\textwidth}
                \centering
                \includegraphics[width=\textwidth]{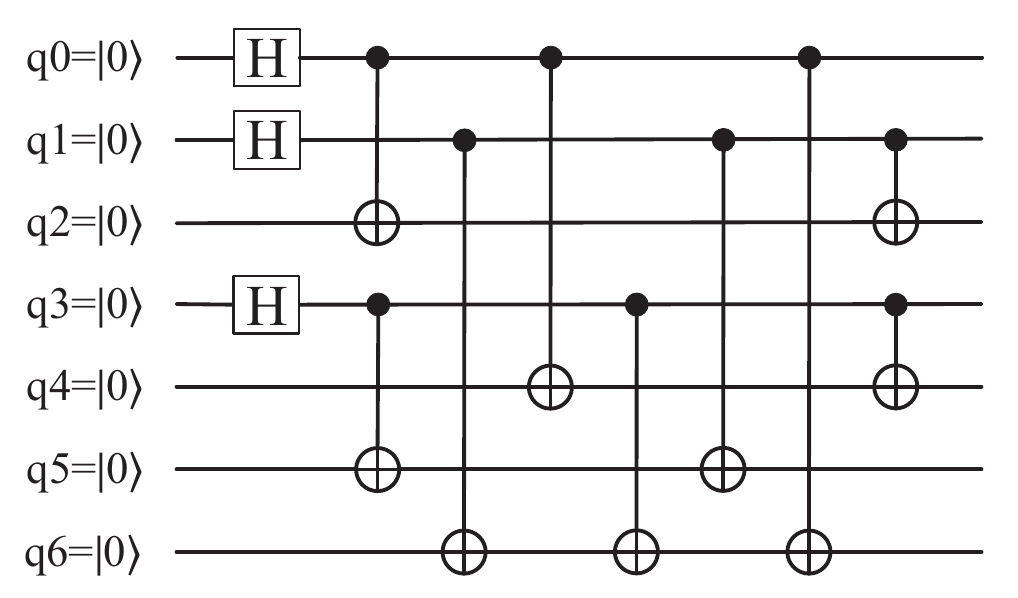}
                \caption{}
                \label{fig:fig2a}
        \end{subfigure}%
        ~ 
        \begin{subfigure}[b]{0.2\textwidth}
                \centering
                \begin{tabular}{|r|l|}\hline
                \# & Instructions\\ \hline & QUBIT  q0, 0\\ & QUBIT  q1, 0\\ & QUBIT  q2, 0\\ & QUBIT  q3, 0\\ & QUBIT  q4, 0\\ & QUBIT  q5, 0\\ & QUBIT  q6, 0\\
                1 & H  q0\\2 & H  q1\\3 & H  q3\\4 & CNOT  q2, q0\\5 & CNOT  q5, q3\\6 & CNOT  q6, q1\\7 & CNOT  q4, q0\\8 & CNOT  q6, q3\\9 & CNOT  q5, q1\\10 & 
                CNOT  q6, q0\\11 & CNOT  q2, q1\\12 & CNOT  q4, q3\\\hline
                \end{tabular}
                \caption{}
                \label{fig:fig2b}
        \end{subfigure}
        \caption{(a) Circuit for preparing the logical 0 using Steane Code [[7,1,3]], (b) QASM program description of the logical zero preparation circuit}
\end{figure}

As an example of a quantum circuit, fig. \ref{fig:fig2a} shows a circuit for preparing logical zero using Steane Code [[7,1,3]] (the output is shown as $\Ket{\overline{0}}$). Note that in this circuit, qubits are initialized to $\Ket{0}$. Each qubit participates in several one- and two-qubit operations. Seven physical qubits at the output of the circuit represent a bundled logical qubit. Fig. \ref{fig:fig2b} shows the QASM description of this encoding circuit. The first part of the QASM file describes the complete set of qubits involved in the circuit and their initial state, e.g. $\Ket{0}$, $\Ket{1}$, $\Ket{+}$ or $\Ket{-}$. The second part of the QASM file describes the quantum operations based on their execution order. Operations having no dependency, e.g. H q0 and H q1, can be added to the file with any ordering. In each line, the operation type, e.g. CNOT, is followed by the qubit(s) involved in the operation. If two qubits are involved in the operation, the first mentioned qubit determines the \textit{control qubit} and the second one determines the \textit{target qubit}. The QIDG for the circuit is built from this QASM description after applying quantum fabric synthesis.

\subsection{ Trapped-ion technology}

Trapped-ion technology is a promising technology for implementing quantum circuits \cite{ladd_quantum_2010}. Hence, it is selected as the underlying technology for our proposed quantum CAD tool. However, our tool can in principle handle other technologies such as the superconducting and solid state quantum fabrics. A recent specification of the trapped-ion technology emanates from a \textit{government furnished information} (GFI) document released by the Sandia National Laboratories. A synopsis of relevant details and the layout architecture for this new trapped-ion quantum fabric is included in the remainder of this section.

\textbf{Qubits:} Qubits are realized by ions. The ions can be moved by applying electrical field.

\textbf{Wells and movement channels:} Wells are connected together through movement channels. A well is called \textit{free}, if no qubit occupies it. Two wells are called \textit{adjacent}, if they are connected to each other through a movement channel. There are three types of wells:

\begin{enumerate}
\item  \textbf{Basic Well:} This well only supports qubit movement.\textbf{}

\item \textbf{Creation Well:} The qubits are fed to the fabric through creation wells. Besides, it allows qubits to move through it. \textbf{}

\item \textbf{Interaction Well:} Quantum operations (also called instructions in this paper) are performed in interaction wells. These wells also allow qubit movement. During a quantum operation, the well is reserved and does not allow another operation or movement. The state-of-the-art trapped-ion technology supports quantum operations involving up to two qubits \cite{rowe_transport_2002}. In a one-qubit operation, only the target qubit resides in a well. In a two-qubit operation, two qubits inhabit a well, control qubit remains untouched while the target qubit is affected by the operation.\footnote{ There are cases in which both qubits are affected. For the sake of simplicity, we still call one qubit as control qubit and the other one as target qubit.}
\end{enumerate}

\begin{figure}
\centering
\scalebox{0.3}{\includegraphics[width=3\textwidth]{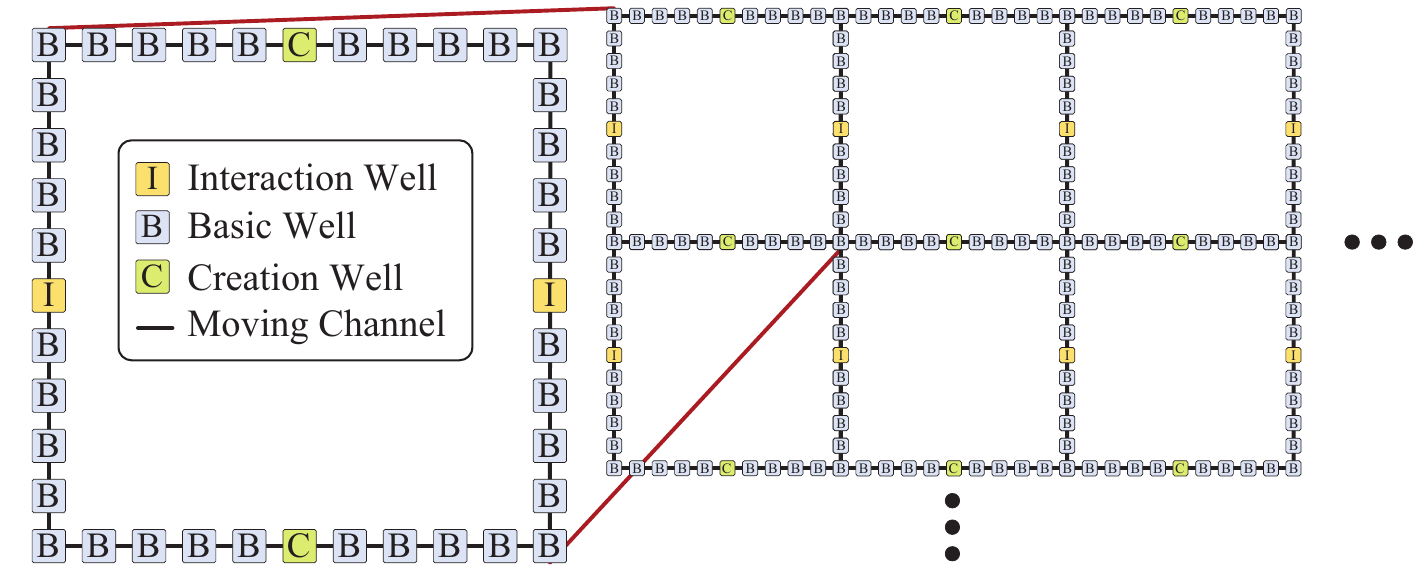}}
\caption{A model of trapped-ion fabric}
\label{fig:fig3}
\end{figure}

Fig. \ref{fig:fig3} shows the trapped-ion circuit fabric model. Wells are represented by squares while moving channels are drawn as line segments. White spaces in this figure represent empty locations on the fabric. As can be seen, this fabric consists of 11$\times$11 templates. Each ULB shown in fig. \ref{fig:fig1} is an n$\times$n grid of these templates. Determining how to optimize the size of ULB is discussed in the section \ref{cadflow}.

Qubits travel through the wells and channels to reach interaction wells for quantum operations. There are three constraints for qubit relocations: 1) Movement is allowed only between two adjacent wells, 2) During qubit creation or interaction, the corresponding well is reserved and does not allow another operation or movement, and 3) Moving channels are \textit{half-duplex}.\textbf{}

\textit{Well capacity} can be defined as the maximum number of qubits that can concurrently reside inside a well. References \cite{rowe_transport_2002}, \cite{hensinger_t-junction_2006}, and \cite{leibrandt_demonstration_2009} discuss about the practical details of how more than one qubit inhabit a channel. Well capacity is a technology-specific parameter. In this paper, the well capacity is set to five according to the GFI document.

\section{ The CAD flow for quantum circuits in the trapped-ion circuit fabric}
\label{cadflow}

The input of QUFD is a QASM file describing the FT implementation of the quantum operation (including the required syndrome extraction circuit) in terms of native quantum instructions. The output of QUFD is a list of \textit{low-level control commands}. These commands have three types: 
\begin{enumerate}
\item Qubit movement command from a source well to a destination well
\item Qubit creation command at a creation well
\item Operation command on one or two qubits at an interaction well
\end{enumerate}

\begin{figure}
\centering
\scalebox{0.5}{\includegraphics*[width=1.8\textwidth]{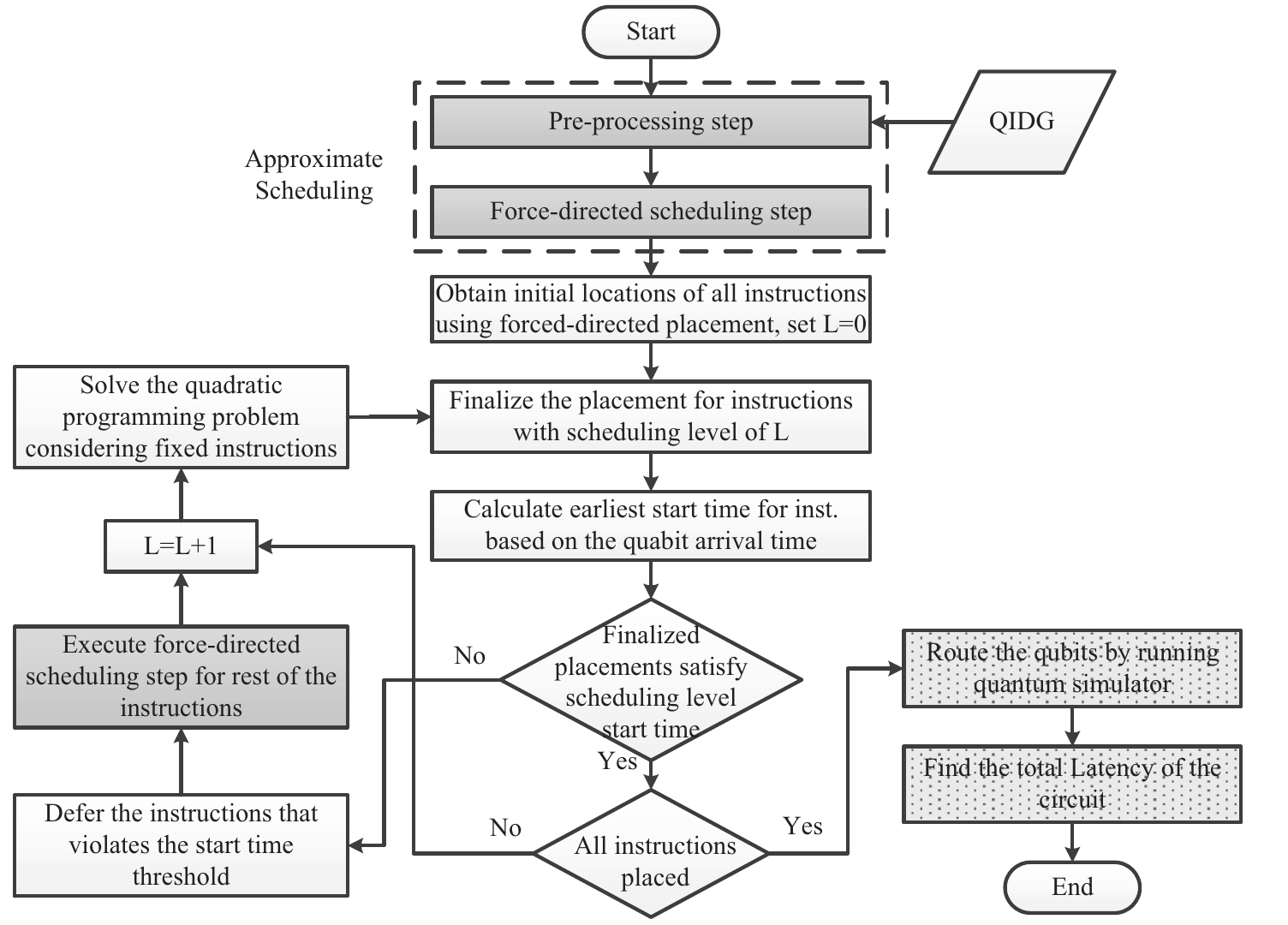}}
\caption{Flow of the Quantum physical mapper}
\label{fig:fig4}
\end{figure}

The QUFD CAD flow for mapping the circuit description of a target FT quantum operation to a ULB is shown in fig. \ref{fig:fig4}. The first step in the flow is to make a QIDG from the QASM file. QIDG is built based on the partial ordering of native instructions (which we shall again call quantum `instructions' for simplicity). The scheduling, placement and routing steps of the quantum physical mapper are highly dependent to one another. More precisely, the scheduling solution affects the quality of the placement solution due to the resource contention, whereas placement solution changes the scheduling solution due to the routing latency. Similarly, the routing solution changes the scheduling solution and affects the placement solution indirectly. The proposed flow for the quantum physical mapper captures these dependencies by applying a scheduling step, a timing-driven instruction placement that minimizes the estimated routing latencies with iterative calls to the force-directed scheduling step to re-calculate  scheduling levels for the instructions, and a routing step that determines the exact qubit routes based on the provided placement solution. The goal of the routing step is to minimize the additional routing latency on the critical path(s) of the circuit. 

Routing is done in two steps: \textit{static routing} and \textit{dynamic routing}. In the static routing step, the shortest route for each qubit movement is determined. In the dynamic routing step, the actual control commands for routing the qubits are generated. In this step, the router may prohibit a qubit from entering a routing channel, if the channel is full. In this case, the qubit in question should wait till the channel usage drops below its capacity. Details of the different steps in the mapper flow (except for the router which is quite straightforward) are described in the following sections.

In addition to minimizing the latency of an FT quantum operation on a given ULB, QUFD minimizes the expected latency of the quantum algorithm by exploring different ULB sizes the determine the best size. The expected latency of the quantum algorithm is defined as a weighted sum of latencies of quantum circuit realizations of the FT quantum operations in the ULB plus the average inter-ULB routing latency. Analyzing the (timing) critical subgraph of the QODG, one can find the occurrence frequency of each FT operation ($w^o$). Latency of each FT operation as a function of ULB size is found by applying QUFD to their QIDG. These latencies are denoted by $L^o(n)$ in an n$\times$n ULB. Moreover, the average inter-ULB routing distance ($D^{r,avg}$), which is the average qubit routing latency in the QODG critical path normalized to the routing latency between two adjacent ULBs, is found by mapping the QODG to the quantum circuit fabric. The routing latency between adjacent ULBs is denoted by $L^{r,1}$ for 1$\times$1 ULB and is linearly proportional to the ULB size. The best ULB size for the target quantum algorithm is found by the following equation:

\begin{equation}
{n^{best}} = \mathop {{\text{argmin}}}\limits_n \left( {n{L^{r,1}}{D^{r,avg}} + \mathop \sum \limits_{o \in FT \; oper. \; set} {w^o}{L^o}(n)} \right) \label{formula1}
\end{equation}

\begin{figure}
\centering
\scalebox{1.0}{\includegraphics*[width=0.7\textwidth]{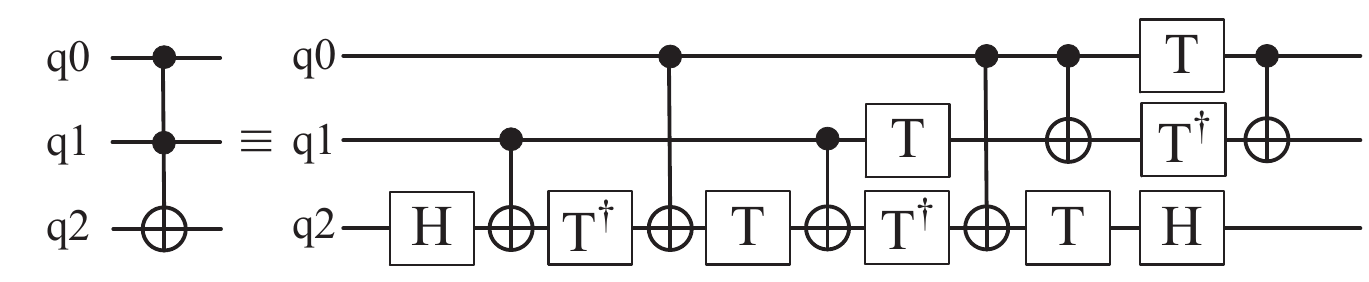}}
\caption{Fault-tolerant implementation of Toffoli circuit}
\label{fig:Toffoli}
\end{figure}

To illustrate the process of setting up equation \eqref{formula1} for a QODG, we consider a Toffoli gate with two control qubits. The Toffoli gate can be implemented using 15 FT operations \cite{shende_cnot-cost_2009} shown in fig. \ref{fig:Toffoli}. Now, consider a 2$\times$2 mesh TQA. ULBs in this architecture are numbered from 1 to 4 (1 for top-left , 2 for top-right, 3 for bottom-left and 4 for bottom-right ULB). A QODG mapping tool can map the Toffoli circuit to this architecture. An exemplary (high quality) mapping solution for this circuit is depicted in table \ref{table:toffolitable}.   

\begin{table}
\centering
\caption{Placement solution and cost for fault-tolerant operations in Toffoli circuit shown in fig. \ref{fig:Toffoli}}
\begin{tabular}{|c|l|c|c|c|c|c|c|c|c|l|}
\hline
& & & \multicolumn{3}{|c|}{Location at beg.} & \multicolumn{3}{|c|}{Location at end} & \\
\cline{4-9}
\ Schedule & Instruction & Placement & q0 & q1 & q2 & q0 & q1 & q2 & Cost \\ 
\hline 
1 & H  q2 & 3 & 1 & 2 & 3 & 1 & 2 & 3 & $L^H(n)$\\\hline 
2 & CNOT  q1,q2 & 2 & 1 & 2 & 3 & 1 & 2 & 2 & $L^{CNOT}(n)+2nL^{r,1}$\\\hline 
3 & $T^{\dagger}$  q2 & 4 & 1 & 2 & 2 & 1 & 2 & 4 & $L^{T^{\dagger}}(n)+nL^{r,1}$\\\hline 
4 & CNOT  q0,q2 & 1 & 1 & 2 & 4 & 1 & 2 & 1 & $L^{CNOT}(n)+2nL^{r,1}$\\\hline 
5 & T  q2 & 3 & 1 & 2 & 1 & 1 & 2 & 3 & $L^{T}(n)+nL^{r,1}$\\\hline 
6 & CNOT  q1,q2 & 2 & 1 & 2 & 3 & 1 & 2 & 2 & $L^{CNOT}(n)+2nL^{r,1}$\\\hline 
7.a & $T^{\dagger}$  q2 & 4 & 1 & 2 & 2 & 1 & 2 & 4 & $L^{T^{\dagger}}(n)+nL^{r,1}$\\\hline 
7.b & T  q1 & 2 & 1 & 2 & 2 & 1 & 2 & 4 & $L^{T}(n)$\\\hline 
8 & CNOT  q0,q2 & 4 & 1 & 2 & 4 & 4 & 2 & 4 & $L^{CNOT}(n)+nL^{r,1}$\\\hline 
9.a & T  q2 & 3 & 4 & 2 & 4 & 2 & 2 & 3 & $L^{T}(n)+nL^{r,1}$\\\hline 
9.b & CNOT  q0,q1 & 4 & 4 & 2 & 4 & 2 & 2 & 3 & $L^{CNOT}(n)+nL^{r,1}$\\\hline 
10.a & H  q2 & 3 & 2 & 2 & 3 & 1 & 2 & 3 & $L^{H}(n)$\\\hline 
10.b & T  q0 & 1 & 2 & 2 & 3 & 1 & 2 & 3 & $L^{T}(n)+nL^{r,1}$\\\hline 
10.c & $T^{\dagger}$  q1 & 2 & 2 & 2 & 3 & 1 & 2 & 3 & $L^{T^{\dagger}}(n)$\\\hline 
11 & CNOT  q0,q1 & 1 & 1 & 2 & 3 & 1 & 1 & 3 & $L^{CNOT}(n)+nL^{r,1}$\\
\hline
\end{tabular}
\label{table:toffolitable}
\end{table}

Note that operations 7.b, 9.a, 10.a and 10.c are not part of the critical path of this circuit. As can be seen, the total latency of the critical path of the circuit is $2\times L^T(n) + 2 \times L^{T^{\dagger}}(n) + 6 \times L^{CNOT}(n) +1 \times L^H(n) + 13 \times nL^{r,1}$. In section \ref{results}, after finding the latency for each FT operation as a function of the ULB size, the value of the Toffoli gate latency is reported as a function $n$ and the best ULB size is chosen.

\begin{figure}
\centering
\scalebox{0.5}{\includegraphics*[width=1.8\textwidth]{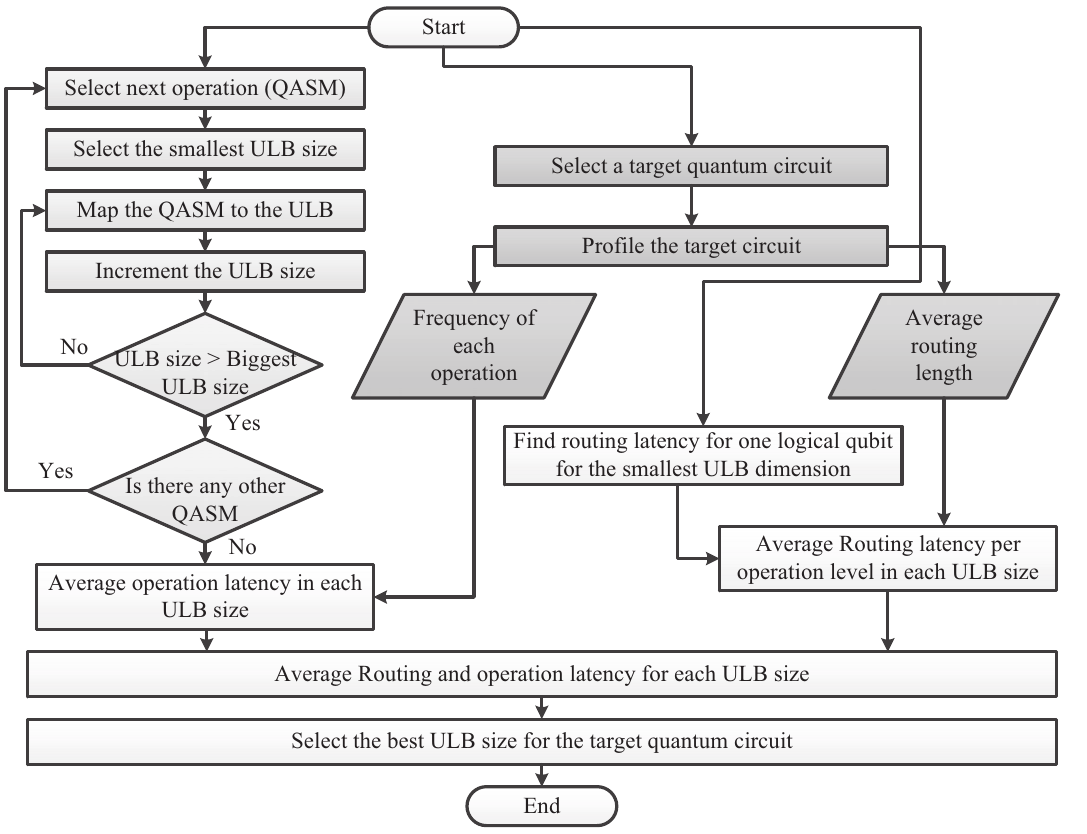}}
\caption{ULB sizing flow}
\label{fig:fig5}
\end{figure}

Fig. \ref{fig:fig5} shows the QUFD ULB sizing flow. In this flow, the smallest ULB size is set by the size required for the most complex operation circuit that needs to be mapped to the quantum fabric (this is typically a non-transversal gate operation). As the ULB size increases, the resource contention decreases (i.e., there is less routing resource contention and thus potentially shorter waiting times to access routing channels during the dynamic routing step); however, the intra-ULB routes become longer, and hence, routing latencies may actually become higher . From another perspective, increasing the ULB size increases the inter-ULB routing latency. In practice, there is an optimum ULB size that minimizes the aforesaid cost function in \ref{formula1}. 

The proposed ULB sizing flow is an offline process to generate the tiles that comprise the TQA and the gate library that is used by the quantum physical mapper. Considering this fact, long run-time for this process does not raise a problem. Moreover, the number of ULB sizes explored to find the optimal $n$ is finite considering the fact that beyond a certain ULB size, increasing the ULB size does not positively affect the mapped circuit latency. 

\section{ Quantum physical mapping problem}
\label{physicalmapping}

The quantum physical mapper converts a quantum circuit to a list of low-level control commands by specifying the exact qubit movements and instruction execution times. Quantum physical mapping problem includes scheduling, placement and routing sub-problems. The scheduling step determines the order of instruction execution, the placement step determines locations of qubits at the beginning and end of each operation as well as the location for each instruction execution, while the routing step determines the qubit location at movement times.

Qubit movement latency is a big part of the quantum circuit latency. Therefore, the routing latency (which is a function of instruction placement and channel congestion) must be included in the decision making procedure for scheduling and placement problems. To account for routing latency in scheduling and placement steps, routing latency estimates based on the physical distance between interaction wells are used. Physical distance is different from Manhattan distance because it is the shortest path between two wells going through movement channels. Hence, it is equal or longer than Manhattan distance.

To explain the concurrent scheduling and placement problem in quantum physical mapper and the similarities and differences between this problem and scheduling and placement problem in the traditional VLSI CAD field, we consider two scenarios. In the first scenario, each instruction is executed in a unique interaction well and in the second scenario, more than one instruction can be executed in one interaction well as long as their execution times do not overlap. 

\textbf{First Scenario}: Interaction wells in this scenario are similar to gates or transistors in VLSI circuits with the difference that there are shared connectors (channels) between interaction wells in the quantum fabric in place of dedicated connectors (point-to-point wires) in the VLSI circuits. Considering the large channel capacity in the quantum fabric, we can ignore the effect of routing resource contention in this scenario. Under this assumption, the best scheduling policy is the \textit{as soon as possible schedule}. Moreover, the placement problem in this scenario is similar to VLSI timing-driven placement problems. The goal of the optimization is to minimize the total added latency to the critical path of the quantum circuit.

\textbf{Second Scenario}: Sharing interaction wells between instructions complicates the scheduling and placement problems. The scheduling and placement problems can thus be seen as a 3-dimensional placement problem with time as the third dimension. In addition to the constraint of no-overlap between execution times of instructions placed at the same interaction well, placing instructions along the time axis must satisfy the instruction dependency constraints. The resulting circuit latency and area in this scenario are much smaller than those of the first scenario.

\begin{figure}
        \centering
        \begin{subfigure}[b]{\textwidth}
                \centering
                \includegraphics[width=0.6\textwidth]{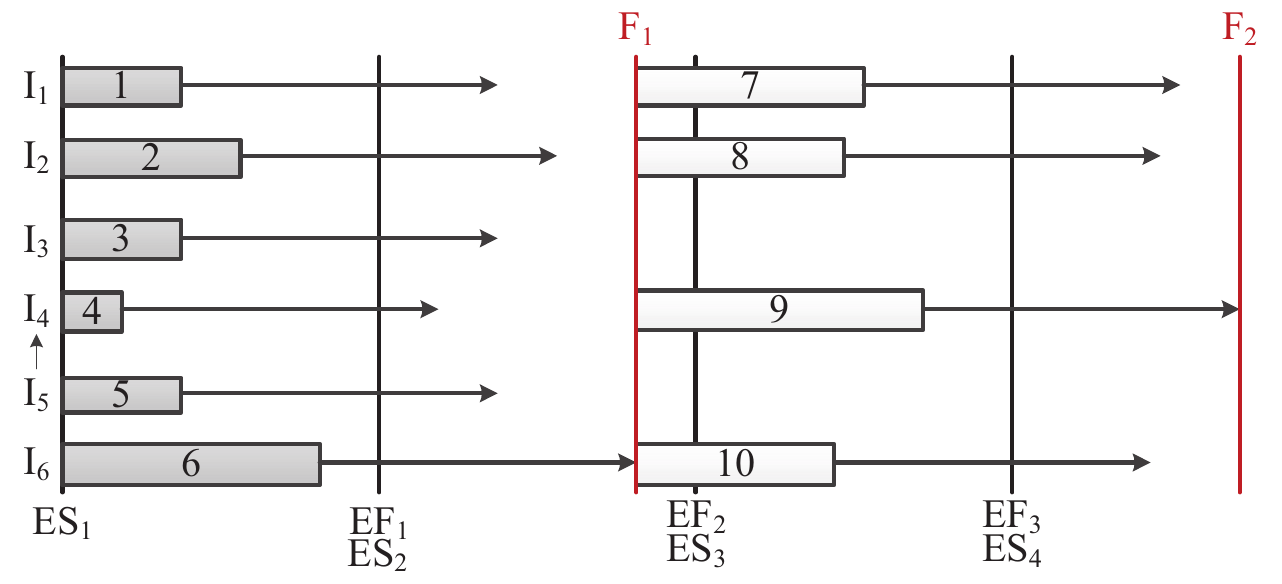}
                \caption{}
                \label{fig:fig6a}
        \end{subfigure}%
        
        \begin{subfigure}[b]{\textwidth}
                \centering
                \includegraphics[width=0.6\textwidth]{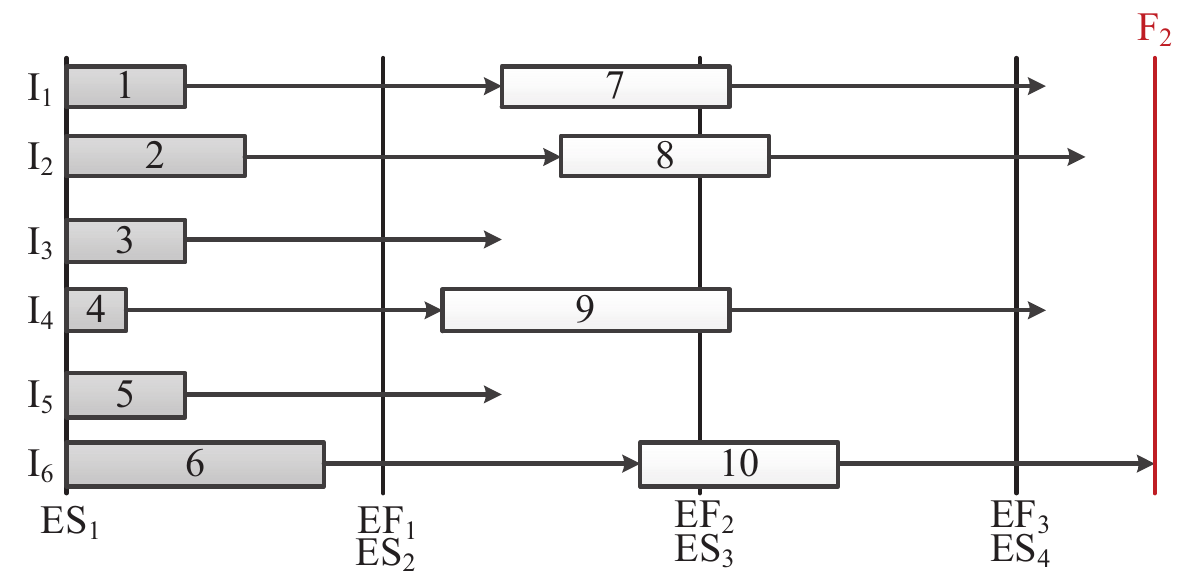}
                \caption{}
                \label{fig:fig6b}
        \end{subfigure}
        \caption{Example of scheduling and placement solutions: (a) slotted time with fixed start time, (b) Slotted time with ASAP start time}
\end{figure}

To simplify the placement problem, we consider discrete start times for instruction executions. An example of the scheduling and placement solution for this scenario is shown in fig. \ref{fig:fig6a}, where \textit{ES${}_{t}$}, \textit{EF${}_{t}$} and \textit{F${}_{t}$} denote the expected start time, expected finish time, and actual finish time of the instructions in time slot \textit{t}, respectively. Moreover, \textit{I${}_{1}$} to \textit{I${}_{6}$} indicate the interaction wells on the quantum fabric; numbers in rectangles show the routing times of the operand qubits, and horizontal arrows represent the execution latencies of instructions. Instructions with dark color belong to the first scheduling level whereas instructions with white background belong to the second scheduling level.

\begin{figure}
        \centering
        \begin{subfigure}[b]{\textwidth}
                \centering
                \includegraphics[width=0.6\textwidth]{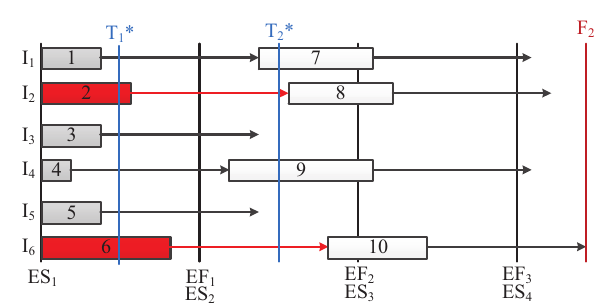}
                \caption{}
                \label{fig:fig7a}
        \end{subfigure}%

        \begin{subfigure}[b]{\textwidth}
                \centering
                \includegraphics[width=0.6\textwidth]{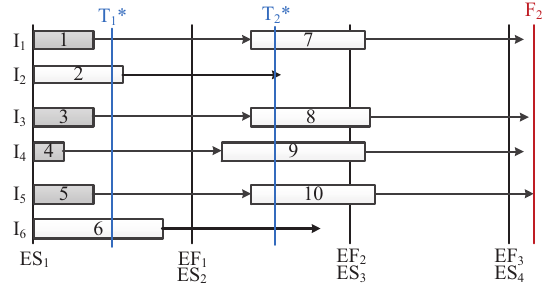}
                \caption{}
                \label{fig:fig7b}
        \end{subfigure}
        \caption{Example of the proposed method for modifying the scheduling and placement solutions: (a) before modification, (b) after modification}
\end{figure}

As can be seen in fig. \ref{fig:fig6a}, considering synchronized start time for qubit routing in each scheduling level (\textit{ES${}_{1}$} for level 1 and \textit{F${}_{1}$} for level 2) leads to inefficient use of resources and increases the circuit latency. To solve this problem, if instructions are started as soon as their qubits and the host interaction well become ready, the total latency of the circuit is lowered as shown in fig. \ref{fig:fig6b}. This scheme has commonly been used in previous work such as \cite{dousti_minimizing_2012}. Unfortunately, this modification decreases the  predictability of start times for instruction executions. More precisely, this scheduling contradicts the assumption about separation of scheduling and placement steps because it makes it possible for two instructions that have statically been scheduled in two different time slots to overlap in the actual schedule of the circuit due to changes to routing latencies. Furthermore, placing instructions on the circuit fabric based on the instruction level has an underlying assumption about the stop time of the previous-level instructions. This assumption does not hold if the instructions are scheduled as soon as their qubits and interaction wells are ready. An example of this case is shown in fig. \ref{fig:fig7a}. As can be seen, the second scheduling level instructions that are placed in the interaction wells hosting instructions 2 and 6 in the first scheduling level ($I_{2}$ and $I_{6}$) are deferred. If the placement module had been aware of this situation, instructions 8 and 10 would have been placed on other interaction wells to lower their finish times. 

To be able to predict the actual instruction execution times in the scheduling and placement step, our approach calculates the estimated start time of each instruction and if the estimated start time exceeds a start time threshold (shown by $T^*_i\ $in fig. \ref{fig:fig7a} and calculated as the middle point of ${ES}_{i}$ and ${EF}_{i}$), the instruction will be deferred and the scheduling and placement solutions are modified. An example of this modification is shown fig. \ref{fig:fig7b}. As can be seen, this modification reduces the total latency of these ten instructions. Details of the proposed scheduling and placement solutions are presented in sections \ref{sched} and \ref{placement}, respectively.

\section{ Instruction scheduling}
\label{sched}

The first step of the quantum physical mapper is \textit{loose scheduling}. The goal of this step is to assign each instruction to a scheduling level (or time slot) to minimize the total latency of the circuit subject to dependency and resource availability constraints. All instructions assigned to one scheduling level can be executed simultaneously. This step is called loose because it does not consider the routing latencies when scheduling the instructions. To modify the scheduling solution based on placement decisions, a part of this scheduling step is iteratively called during the placement procedure.

The input to the scheduling procedure is a QIDG, which is a \textit{directed acyclic graph}, capturing the dependencies between instructions specified in the QASM file. Each node of this graph is a native quantum instruction, i.e., it can be performed in one step in an interaction well in the trapped-ion quantum circuit fabric. Two instructions are dependent if any of these conditions hold:

\begin{enumerate}
\item  The target qubit of instruction A is used as the control qubit of instruction B (Read After Write (RAW) dependency)

\item  The target qubit of instruction A is used as the target qubit of instruction B (Write After Write (WAW) dependency)

\item  The control qubit of instruction A is used as the target qubit of instruction B (Write After Read (WAR) dependency).
\end{enumerate}

\begin{figure}[h]
\centering
\includegraphics*[width=0.4\textwidth]{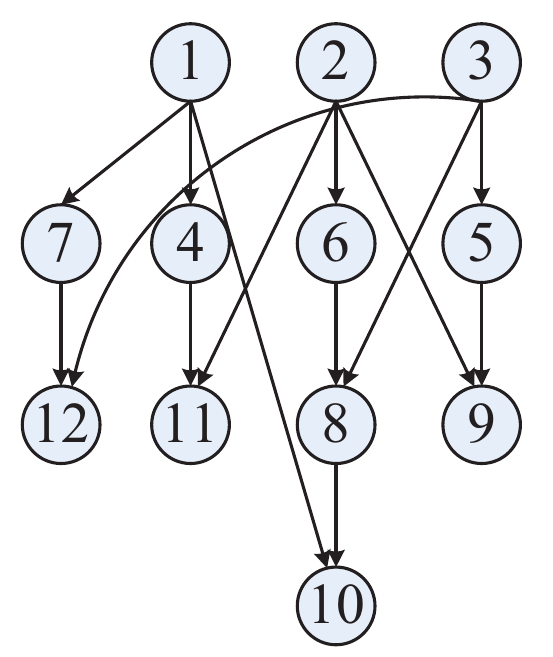}
\caption{Example of QIDG for the quantum circuit presented in fig. \ref{fig:fig2a}}
\label{fig:fig8}
\end{figure}

Each dependency is shown using a directed edge from the vertex representing instruction A to the vertex representing instruction B. Fig. \ref{fig:fig8} shows an example of QIDG for the quantum circuit presented in fig. \ref{fig:fig2a}. 

Each instruction and each qubit in the QIDG are identified by unique identifiers, represented by index $i$ and $q$ respectively. In the QIDG, nodes with the same parent instruction are called \textit{sibling instructions}. These instructions are two-qubit instructions that share the same control qubit. 

Due to the \textit{no-cloning theorem} \cite{wootters_single_1982}, only one copy of each qubit can exist in the circuit. More precisely, only one instruction can access a qubit at each point in time. So, if one qubit needs to be accessed by sibling instructions, e.g. qubit q0 in instructions 4 and 7 in fig. \ref{fig:fig8}, those instructions cannot be executed at the same time. Based on the assumption that all instructions in one scheduling level can be executed simultaneously, at most one instruction from each sibling instruction set can be placed in one scheduling level. Therefore, an ordering should be imposed among sibling instructions to be able to schedule the instructions without additional restrictions. 

To order the sibling instructions, we add \textit{auxiliary edges} between them. Note that adding these auxiliary edges may increase the critical path length. Hence, this process should be done carefully. From another perspective, the number of instructions at each scheduling level is limited by the number of available interaction wells in the ULB.

The minimum latency QIDG scheduling problem (called \textit{quantum instruction scheduling problem} or QISP) with the assumption of the same execution length for every instruction may be formulated as follows:

\begin{align}
\intertext{
\begin{center}
$Min \; L$
\end{center} 
\noindent subject to:}
& t_i\le L, & \forall i \label{qisp1} \\
& t_i=\sum\nolimits_{l}{l.x_{il}}\, & \label{qisp2}\\
& \sum\nolimits_{l}{x_{il}}=1, & x_{il}\in \{0,1\} \label{qisp3} \\
& t_i-t_j\ge 1, & \ \forall j\in P_i \label{qisp4} \\
& x_{il}+\sum\nolimits_{k\in S_i}{x_{kl}}\le 1, & \forall l,i \label{qisp5} \\
& \sum\nolimits_{i}{x_{il}}\le N^{max\ }, & \forall l \label{qisp6}
\end{align}

\noindent where $P_i$ and $S_i$ denote the sets of \textit{parent} and \textit{sibling instructions} for instruction $i$, respectively;  $t_i$ denotes the start time of instruction $i$; \textit{L}, which stands for latency, is defined as the time when the last instruction starts its execution; $N^{max}$ denotes the maximum number of concurrent instructions at each time. Parameter $x_{il}$ is the scheduling parameter, which is equal to one if instruction $i$ starts from level \textit{l }; otherwise it is\textit{ }zero. Equation \eqref{qisp2} relates the start time of an instruction to the scheduling parameter ($x_{il}$). Constraint \eqref{qisp3} forces an instruction to have only one starting time. Constraint \eqref{qisp4} captures the dependency constraints in the QIDG. Constraint \eqref{qisp5} captures the fact that sibling instructions cannot be scheduled at the same time. Constraint \eqref{qisp6} enforces the limit on the number of concurrent instructions at each scheduling level. 

QISP is an integer linear programming (ILP) problem. In fact, QISP is a \textit{resource-constrained scheduling} problem \cite{demicheli_synthesis_1994} with one important difference. In QISP, in addition to the limit on the total number of processing elements in each scheduling level (cf. constraint \eqref{qisp6}), the number of processing elements that can be used for sibling instructions is limited to one (cf. constraint \eqref{qisp5}). Our solution to QISP decomposes the problem into two steps: (i) A pre-processing step based on the \textit{list scheduling} algorithm \cite{davidson_experiments_1981} is applied to the QIDG to \textit{totally order} the sibling instructions; (ii) An enhanced version of the well-known \textit{force-directed scheduling} algorithm \cite{paulin_force-directed_1989} is applied to the modified QIDG so as to minimize a cost function which is a function of the number of scheduling levels and the number of concurrent instructions per scheduling level. Details are provided below.  

\subsection{ QIDG pre-processing step}

The goal of the pre-processing step is to impose a total ordering on the sibling instructions to avoid any violations of constraint \eqref{qisp5} during the scheduling process. A well-known approach for solving the resource-constrained scheduling problem is the list scheduling algorithm \cite{davidson_experiments_1981}, which is reviewed in the next paragraph. 

First,\textit{ as soon as possible} (ASAP) and \textit{as late as possible} (ALAP) schedules for a given task dependency graph are constructed. The difference between levels of an instruction in the ASAP and ALAP schedules, which is called \textit{mobility}, sets the priority of an instruction in scheduling with respect to the other instructions (that is, less mobile instructions have higher priority to be scheduled first). To schedule the instructions based on this approach, starting from the first level of instructions in the ASAP schedule, if a resource constraint is violated, an instruction with the highest mobility (least priority) will be picked and delayed by one scheduling level. While the resource constraint is violated, this process continues. This process is repeated for the next level and continues until all conflicts are resolved. The motivation behind this greedy heuristic is that the instruction with higher mobility can be delayed without adversely affecting the overall execution latency.

Algorithm \ref{alg:pre} provides the pseudo-code for the pre-processing step of the scheduling solution, which is based on list scheduling algorithm. One of the difficulties in QISP is that the length of the critical path that determines the instruction mobilities is not known \textit{a priori} because the sibling instructions have not been ordered yet. Therefore, new ASAP and ALAP schedules are determined in each step based on the modified QIDG (namely, current partial ordering of the sibling instructions). Let ${SL}_i$, $ASAP_i$, and $ALAP_i$ denote the scheduling level, ASAP level, and ALAP level of instruction \textit{i}. ASAP and ALAP levels of each instruction along with the priority values (i.e., $1/(ALAP_i-ASAP_i)$) are calculated in the first iteration. Starting from the set of instructions with $ASAP_i=0$, an instruction with the highest priority among siblings that have an ASAP level equal to zero is selected. This instruction is added as a parent for its sibling instructions with the same or higher ASAP level. Note that this selection fixes the instruction with the least mobility and moves its sibling instructions to the next levels in contrast to the list scheduling that repeatedly defers the instruction with the highest mobility until the resource constraint is satisfied. In this special problem, these approaches have the same results (with lower complexity for our algorithm), because the number of allowed sibling instructions in each scheduling level is only one. After these modifications, ASAP and ALAP and priority metrics are recalculated. If constraint \eqref{qisp5} is no longer violated for the selected set of instructions, the set of instructions with $ASAP_i=1$ will be selected and the process of adding edges will be repeated. This process is continued for all the ASAP levels or until all required edges are added to the QIDG.

\begin{algorithm}
\caption{Preprocessing of QIDG}
\label{alg:pre}
\begin{algorithmic}[1]
\STATE Compute $ASAP_i$, $ALAP_i$ and $m_i=1/(ALAP_i-ASAP_i)$ \\
\STATE $S_i^\prime=\{j|j\in S_i$ and $ASAP_j=ASAP_i\}$  ,  $A=0$ //target ASAP
	\WHILE {$\cup S_i \ne \O$}
		\REPEAT
			\STATE $flag =false$
			\FOR {$i \in I$}
				\IF{$ASAP_i=A$ and $S_i \ne \emptyset$}
					\STATE ${i^*} = \mathop {argmax}\nolimits_{i \in S_i^\prime } ({m_i})$
					\STATE ${S_{{i^*}}} = \emptyset $ , $flag=true$
					\FOR {$j \in S_i$}
						\STATE ${P_i} = {P_i}\mathop  \cup \{ {i^*}\} $
						\STATE ${S_j} = {S_j} - \{ {i^*}\} $
					\ENDFOR
					\STATE \textbf{break}
				\ENDIF
				\IF {~$flag$}
					\STATE $A=A+1$;
				\ENDIF
			\ENDFOR
		\UNTIL {$A = \max \left( {ASA{P_i}} \right)$ and $~flag$}
		\STATE Compute $ASAP_i$, $ALAP_i$ and $m_i$ \\
		\STATE $S_i^\prime=\{j|j\in S_i$ and $ASAP_j=ASAP_i\}$
 \ENDWHILE
\end{algorithmic}
\end{algorithm}

\begin{figure}[h]
        \centering
        \begin{subfigure}[b]{0.3\textwidth}
                \centering
                \includegraphics[width=0.9\textwidth]{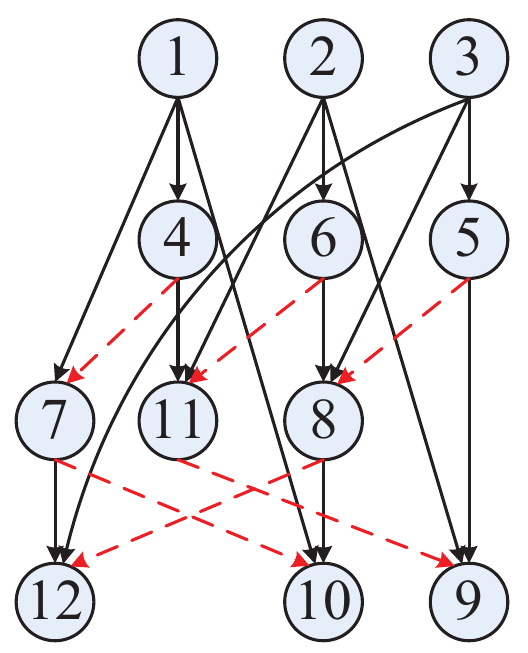}
                \caption{}
                \label{fig:fig9a}
        \end{subfigure}%
        ~ 
        \begin{subfigure}[b]{0.3\textwidth}
                \centering
                \includegraphics[width=0.9\textwidth]{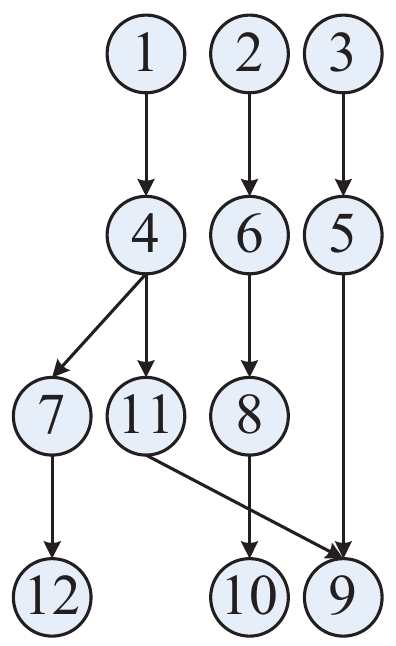}
                \caption{}
                \label{fig:fig9b}
        \end{subfigure}
        \caption{(a) Example of QIDG in fig. \ref{fig:fig8} after preprocessing, (b) Example of scheduled instructions}
\end{figure}

Applying this pre-processing step to the QIDG of fig. \ref{fig:fig8} results in the QIDG of fig. \ref{fig:fig9a}. Dashed (red) edges in this figure show the added auxiliary edges. 

\subsection{ Quantum instruction scheduling algorithm}

The number of quantum instructions being executed at the same time is upper bounded by the number of available interaction wells in the ULB. This bound is captured in constraint \eqref{qisp6} of the QISP formulation. Although increasing the number of concurrent instructions in each scheduling level tends to decrease the number of scheduling levels, it may increase the total latency of the circuit as explained next. Recall that the number of concurrent instructions in each scheduling level greatly affects the placement solution, routing delays, and thus, the latency of the quantum circuit. An instruction cannot be executed until its qubit operands arrive at the host interaction well. The routing time of a qubit is defined as the difference between the arrival time of the qubit in the case of one-qubit instructions (or that of the last arriving qubit in the case of two-qubit instructions) and the time when the instruction was scheduled to be performed. The routing time is a function of the specific route selected for the operand qubit(s) and the wait (stall) time of the qubit during the routing due to the congestion (shared resource contention). Increasing the number of concurrent instructions forces some instructions to be placed in interaction wells that are far from the interaction wells that are hosting their parent instructions. Moreover, increasing the number of concurrent instructions results in an increase in the number of qubits that need to be simultaneously routed. In addition to the effect of longer routes on the routing time, higher number of routed qubits increases congestion, and thus, increases the wait (stall) time. Evidently, increasing the number of concurrent instructions may delay the start of instruction execution times, and thereby, adversely affect the total latency of the quantum circuit. Therefore, to reduce the total latency of a quantum circuit, we must minimize the number of scheduling levels and evenly distribute concurrent instructions among all scheduling levels. This is a multi-objective optimization problem. To solve it, a modified version of the well-known force-directed scheduling (FDS) technique \cite{paulin_force-directed_1989} is used as detailed below.

FDS is a technique used to schedule directed acyclic graphs so as to minimize the resource usage under a latency constraint in a shared resource system. The algorithm consists of three steps: (i) Determine the range of available time slots for scheduling an instruction; (ii) Create a distribution graph, which captures the resource pressures on each time slot, by assuming that an instruction is equally likely to be scheduled to start in any feasible time slot; (iii) A metric called \textit{force} is defined to minimize the resource utilization. By repeatedly assigning instructions to all possible time slots and calculating the force associated with the assignment, several force values will be calculated. The instruction-to-time slot assignment with the lowest force value is chosen, which also balances the instruction concurrencies (i.e., reduces resource pressure).

The first goal of the quantum instruction scheduling is to minimize the number of concurrent instructions at each scheduling level. The second goal is to minimize the number of scheduling levels. To capture both of these goals in our scheduling method, we use a latency-constrained FDS technique which tries to evenly distribute the instructions with fixed circuit latency with the addition of a fixed bound ($N^m$) on the number of concurrent instructions at each scheduling level. For this reason, after assigning an instruction to a scheduling level, the number of  instructions assigned to that scheduling level is incremented by one. When this number reaches $N^m$, no new instruction is assigned to the scheduling level. Based this fact, assignment probabilities in the FDS technique are updated. If an instruction $i$ is found to have no possible scheduling level assignment ($ALAP_i<ASAP_i$), the number of scheduling levels will be incremented by one and the assignment probabilities for the unassigned instructions will be re-calculated. This process continues until all instructions are scheduled. 

Notice that there is a hard global limit on the number of concurrent instructions in each scheduling level, which is $N^{max}$. Our simulation results show that considering a limit less than this global value may decrease the total circuit latency due to a decrease in the routing latency in each scheduling level. Therefore, the scheduling process must be repeated for a few different $N^m \leq N^{max}$ values and the best result (after final routing) chosen. In this paper, five different values for $N^m$ are considered: $N^m\in \left\{N^{ma},0.8N^{ma},0.6N^{ma},0.4N^{ma},0.2N^{ma}\right\}$
\noindent where ${{N}}^{{ma}}{=}{\min \left({{N}}^{{max}},{\mathop{\max }_{{k}} \left|\left\{{i|ASA}{{P}}_{{i}}{=k}\right\}\right|}\right)}$.

\section{ Instruction placement}
\label{placement}

As mentioned in previous sections, the placement solution affects routing latencies, and hence, the start times of instructions. This fact motivates the development of a cross-layer optimization approach spanning the scheduling and placement steps in the quantum physical mapper. The proposed placement algorithm in this section is based on the state-of-the-art placement algorithm for placing logic gates in 2-D layouts. Modifications are made to account for the cross-later optimization issue. The placement solution is an iterative solution, which targets instructions in some scheduling level in each of its iterations. More precisely, at iteration $L$, considering placement solution for instructions in scheduling levels $0$ to $L-1$ has already been obtained, a global instruction placement problem is formulated as a quadratic programming problem and solved by using a quadratic program solver. After this step, a host interaction well is determined for each instruction. Next, the \textit{approximate start time} for each instruction in scheduling level $L$ is calculated based on the fixed placement of its parent instructions and static routing latencies. If an instruction cannot satisfy the start time requirement of scheduling level $L$, it will be deferred by one scheduling level. Then, the force-directed scheduling algorithm is applied to determine the best scheduling solution based on the new deferrals. After applying the scheduling algorithm, the quadratic program solver is applied to the placement problem with new scheduling levels for instructions to obtain (and freeze) the placement solution for instructions in scheduling level $L$. 

In this section, details of this procedure are presented. First, a mathematical formulation of the placement problem is given. Next, a solution without considering the effect of the placement on the scheduling solution is discussed, and, in the end, revisions to the algorithm in order to consider the effect of the routing latencies and placement solution on the scheduling solution are described.

\subsection{ Quantum instruction placement problem formulation}

The ratio of the qubit movement latency to the instruction latency shows the importance of the routing latency in a quantum fabric realization. For example, in fig. \ref{fig:fig3} the minimum Manhattan distance between two interaction wells is 10. Considering 1/10 as the ratio of single move latency to the two-qubit instruction latency according to \cite{whitney_fault_2009}, the minimum routing latency is as large as the latency of performing one instruction, which underlines the importance of the routing latency.

Similar to routing steps, routing latency can be divided into two components: 1) \textit{static routing latency}, and, 2) \textit{dynamic routing latency}. The static routing latency is related to the Manhattan distance between two interaction wells whereas the dynamic routing latency is related to stall times of the qubit in channels due to transient saturation of the channel capacities.

The dynamic behavior of the routing latency is hard to predict in a fully analytical way. In other words, the quantum circuit should in fact be simulated in order to determine the dynamic routing latency. Based on this line of reasoning, only static routing latency is considered in the quantum instruction placement formulation.

There are two types of qubits in the circuit: 1) \textit{I/O qubits}, and, 2) \textit{ancillary qubits}. I/O qubits enter and exit ULBs from specified locations whereas ancillary qubits are created, consumed and then purified to be used again for other instructions. Let $Q^a$, $Q^o$ and $Q$ denote the set of ancillary, I/O  and all qubits in the QIDG, respectively. Similarly, $Q^a_i$, $Q^o_i$ and $Q_i$ denote the set of operand ancillary, I/O and all qubits of instruction $i$, respectively. 

The scheduling level and the parent set for each quantum instruction are determined in the quantum instruction scheduling step. In the quantum instruction placement step, each instruction should be assigned to some interaction well considering its scheduling level. Recall that the placement solution determines the routing latency for each qubit movement, which can change the scheduling level of instructions, and subsequently, affects the start time of the quantum instructions.  

The \textit{quantum instruction placement problem} (QIPP) can be formulated as follows:

\begin{align}
\intertext{
\begin{center}
${min}\left\{\ {\mathop{\max }_{q\in Q^o}\  \left\{t_q\right\}\ \ \ }\right\}$
\end{center} 
\noindent subject to:}
& t_i\ge t_j+d_j+\sum\nolimits_{u\in W}{\sum\nolimits_{v\in W}{y_{iu}y_{jv}R_{uv}}}\, & \forall j\in P_i \label{qipp1} \\
& t_i\ge \sum\nolimits_{c\in C\cup I}{\sum\nolimits_{w\in W}{{y_{iw}z_{qc}I}_{cw}}}\, & \forall q\in Q_i \label{qipp2}\\
& t^f_q\ge t_i+d_i+\sum\nolimits_{c\in I}{\sum\nolimits_{w\in W}{{y_{iw}z_{qc}I}_{cw}}}\ , & \forall i|q\in Q^o_i \label{qipp3} \\
& t_i\ge (t_j+d_j)\sum\nolimits_{w\in W}{y_{iw}y_{jw}}\, & \ if\ \left(t_i\ge t_j\right) \label{qipp4} \\
& y_{iw}\in \{0,1\}, z_{qc}\in \{0,1\}\ & \label{qipp5} \\
& \sum\nolimits_{w\in W}{y_{iw}}=1, \sum\nolimits_{c\in C}{z_{qc}}=1, \forall q\in Q^{a}\label{qipp6}
\end{align}

\noindent where $t_q$ denotes the time when qubit \textit{q} arrives at its output location after finishing the last instruction, $W$, $C$ and $I$ denote the set of locations of the interaction, creation wells and input/output ports in the circuit, respectively, $d_i$ is the latency of the $i^{th}$ instruction, $R_{uv}$ is the static routing latency between the interaction well $u$ and $v$, $R_{cw}$ denotes the static routing time from the creation well (or input port) $c$ to the interaction well $w$, and $y_{iw}$ is an assignment parameter that determines if the instruction is assigned to an interaction well (when it is equal to 1) or not (when 0). Similarly, $z_{qc}$ denotes an assignment parameter that determines if the qubit is assigned to a creation well (or input port) or not. For input qubits, this parameter is fixed. Parameters $y_{iw}$, $z_{qc}$ and $t_i$ are the optimization parameters in the QIPP. Note that the result of the quantum instruction scheduling cannot determine the exact timing of the instructions and can only determine the order of dependent instructions.

The gaol in QIPP is to minimize the total latency in the quantum circuit. Constraint \eqref{qipp1} determines the earliest start time of each instruction based on the start time, static routing time and latency of its parent instructions. Constraint \eqref{qipp2} determines the earliest start time of the quantum instruction that accesses qubit(s) for the first time. Constraint \eqref{qipp3} determines the time when qubits arrive at their output locations after finishing the last instructions. Constraint \eqref{qipp4} limits the number of concurrent instructions that can be executed at each interaction well to one. Constraints \eqref{qipp5} and \eqref{qipp6} define the binary assignment parameters.

\begin{figure}
\centering
\includegraphics*[width=0.7\textwidth]{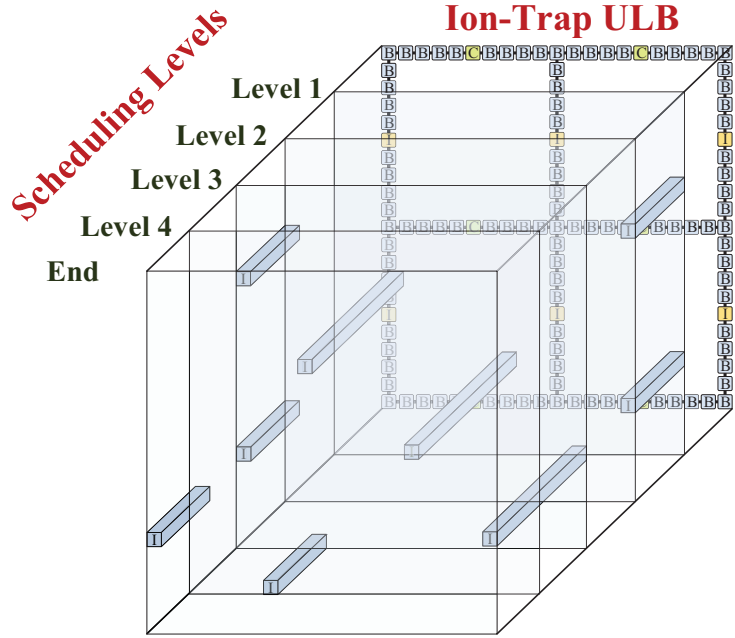}
\caption{Sample instruction placement solution for QIDG of fig. \ref{fig:fig8}}
\label{fig:fig10}
\end{figure}

QIPP is a non-linear, mixed-integer problem. From another perspective, QIPP is a timing-driven 3D placement problem. The first two dimensions correspond to the X and Y dimensions of the quantum fabric whereas the third dimension represents time. In a regular placement problem, the overlap between different placement objects is not acceptable, but in QIPP, the physical overlap is acceptable as long as the execution times of different instructions do not coincide. A sample instruction placement solution for the QIDG shown in fig. \ref{fig:fig9b} is depicted in fig. \ref{fig:fig10}. Each rectangular cube represents an instruction placed at an interaction well on the ULB. Longer rectangular cubes represent two instructions placed at an interaction well in two consecutive scheduling times.

In this paper, to manage the problem complexity, time is discretized. The duration of each time slot (\textit{D}) is considered to be the latency of the fastest instruction such as a rotation operation. Also to solve the timing-driven placement, a net-weighting technique is employed \cite{pan_TimingDriven}. Details of the problem formulation and the algorithm are presented next.

\subsection{ Timing-driven force-directed placement based on fixed instruction levels}

Considering fixed instruction levels based on the result of the quantum instruction scheduling step, the QIPP can be simplified to placing instructions on interaction wells such that there are no overlaps among instructions with the same scheduling level. Note that the routing latency of instructions may delay instructions from their original scheduled time. Thus, minimizing the routing latencies can decrease the time shifts and eventually decreases the total latency of the circuit. However, not all routing latencies have the same importance. An increase in the routing latency of qubits related to an instruction that lies in the critical path of the QIDG will increase the total latency of the circuit, while adding to the routing latency of qubits related to an instruction in a non-critical path may only cause a time shift on the start times of the remaining instructions in that path without changing the total latency of the circuit. To account for this fact, the goal of the placement solution is simplified to minimizing the weighted sum of the routing latencies. The weight for the qubit routing latency between the \textit{i}${}^{th}$ (child) and the \textit{j}${}^{th}$ (parent) instructions is called $m_{ij}$ and is found from equation \eqref{qipp7}.

\begin{equation}
m_{ij}={min}(m^{max},1/(ALAP_i-ASAP_i+{SL}_i-{SL}_j-1)) \label{qipp7}
\end{equation}

\noindent where, $m^{max}$ denotes the maximum weight for cases with infinity weight.

This weight is in inverse proportion to the difference between ALAP and ASAP levels of the child instruction and also the difference between levels of the child and parent instructions. The higher the difference between the child and parent instruction levels, the more time the qubit has to be routed without delaying the child instruction. Moreover, the difference between ALAP and ASAP levels of the child instruction shows the criticality of this instruction to the rest of the circuit.

To model the routing latency in this problem, we use the Manhattan distance between the parent and child instruction locations on the circuit fabric. Using the aforementioned simplifications, the quantum instruction placement problem with \textit{fixed instruction level} (QIPP-FIL) is to minimize the weighted summation of the routing latencies in the quantum circuit. Moreover, no more than one instruction with a specific level can be placed on any interaction well. The effect of initial qubit routing latencies (between the initial qubit locations and the location of the first instructions that use them) and final routing latencies (between the last operation that uses I/O qubits and their final locations) is considered with weights calculated from equation \eqref{qipp7} when ${SL}_i=SL_j+1$. 

QIPP-FIL consists of an instruction placement problem with per-level overlap avoidance constraints plus a qubit placement problem. To solve this problem, an instruction placement technique is used, which is followed by placing the ancillary qubits on the creation sites based on the instruction placement solution.

To place the instructions on the quantum circuit fabric, a modified version of the state-of-the-art force-directed placement tool, SimPL \cite{kim_simpl:_2010}, is used. SimPL works best when placing 10${}^{4}$ to 10${}^{7}$ objects on a circuit fabric. This range complies with the number of instructions that we have for the placement step considering multi-level concatenated error correction scheme for transversal and non-transversal operations. SimPL is reviewed next.

To minimize the half perimeter wire length in VLSI circuits, without any overlap avoidance constraints, an iterative optimization procedure is used. Starting from an initial solution, a quadratic optimization problem is formulated and solved. Distances between placed components are used to formulate a new quadratic optimization problem to be solved. After convergence of the solution, the placement result is the optimal solution. To remove the overlaps, an algorithm called \textit{rough legalization}, which uses a \textit{non-linear scaling} method is proposed in \cite{kim_simpl:_2010}. Based on this approach, over-utilized areas in the fabric are examined and components that are placed in those areas are distributed evenly in the surrounding area. To avoid components from reverting back to their original locations, a \textit{pseudo-net} is added in the new location of any moved component and a wire is added between that pseudo-node and the original component. Weight of the added wire is increased as the number of iterations increases. Note that adding pseudo-nets does not change the problem size because it does not add to the number of variables or equations and only changes the objective function related to the original node. After this modification, the half perimeter wire length minimization problem is formed and solved again. This process continues until all overlaps are removed. Finally, a simple approach is used for final legalization. SimPL improves the speed of prior placement tools and has similar quality to the analytical placement solutions.

Due to the difference between the quantum instruction placement problem and the VLSI placement problem, some modifications must be made to SimPL as explained next. Instead of minimizing the total routing latencies, a weighted summation of the routing latencies is minimized. Moreover, the rough legalization step in this solution is applied to the instructions in each scheduling level. In particular, only overlaps between instructions on the same scheduling level are removed with the non-linear scaling approach. Note that changing the locations of instructions during the rough legalization in one level affects the placement solution for other scheduling levels. Therefore, to consider this effect, the quadratic optimization solution is repeated after finalizing the rough legalization for each scheduling level. To find the final placement solution (assignment of instructions to interaction wells), similar to rough legalization, each scheduling level is processed separately. For each scheduling level, the interaction wells are sorted based on their distance to the instructions, and then the best interaction well for each instruction is chosen. 

After finalizing the placement solution for all instructions, the best creation well for each ancillary qubit is determined. Each ancillary qubit is assigned to the free creation well that is closest to the location of the instruction that uses it for the first time. 

\subsection{ Timing-driven force-directed placement with variable instruction levels}

The proposed method based on classical placement tool plus net-weighting relies on a fixed scheduling level for each instruction. This assumption is violated with the addition of the routing latencies. More precisely, adding the routing latencies may significantly vary the start time of each instruction from its nominal start time. Considering fixed levels for instructions in the placement problem has two negative effects on the placement solution. First, there may be a large number of instructions at some scheduling level, and hence, distributing them on the circuit fabric may cause some of the instructions to be placed far from the interaction wells hosting their parent instructions. However, if the scheduling level of those instructions is changed, they can be placed near the interaction wells hosting their parent instructions. This may even reduce the finish time for the deferred instructions. Second, placing instructions on the circuit fabric based on the fixed scheduling level relies on an assumption about the start and stop times of the previous level instructions. This assumption typically does not, which results in the deferral of every instruction that is placed on the same interaction well in the next scheduling levels, cf. fig. \ref{fig:fig7b}.

In order to avoid big differences between nominal and actual start times for an instruction, a rule can be added to the placement solution as follows. The start time of an instruction assigned to scheduling level $L$ should be less than a threshold time ($T^{th}_{L}$). This threshold start time can be the middle time in time slot $L$. This rule results in higher accuracy of the execution time prediction for instructions in different scheduling levels. To modify the placement solution based on this rule, we propose an iterative approach. In each iteration, instructions in one scheduling level are fixed. In iteration $L$: (i) considering fixed placement solution for instructions in scheduling level $0$ to $L-1$, the proposed solution for QIPP-FIL is applied; (ii) an approximate start time for each instruction (${T}^{min}_i$) in scheduling level $L$ is calculated based on the distance between the location of the interaction well assigned to it and locations of the interaction wells hosting the parent instructions. If an instruction cannot satisfy the start time condition (${T}^{min}_i<=T^{th}_{L}$), it will be deferred by one scheduling level (${SL}^{min}_i=L+1$); (iii) All of the finalized instructions that do not satisfy the timing constraints are removed from the quantum circuit fabric; (iv) The force-directed scheduling step is applied to determine the best scheduling solution for all 'floating' instructions based on the new deferrals. The instruction scheduling step is augmented to support the new parameter (${SL}^{min}_i$); (v) The proposed solution for QIPP-FIL is applied to the placement problem with new scheduling levels for instructions to fix the placement solution for the remaining instructions in the scheduling level $L$.

\begin{figure}
\centering
\includegraphics*[width=0.7\textwidth]{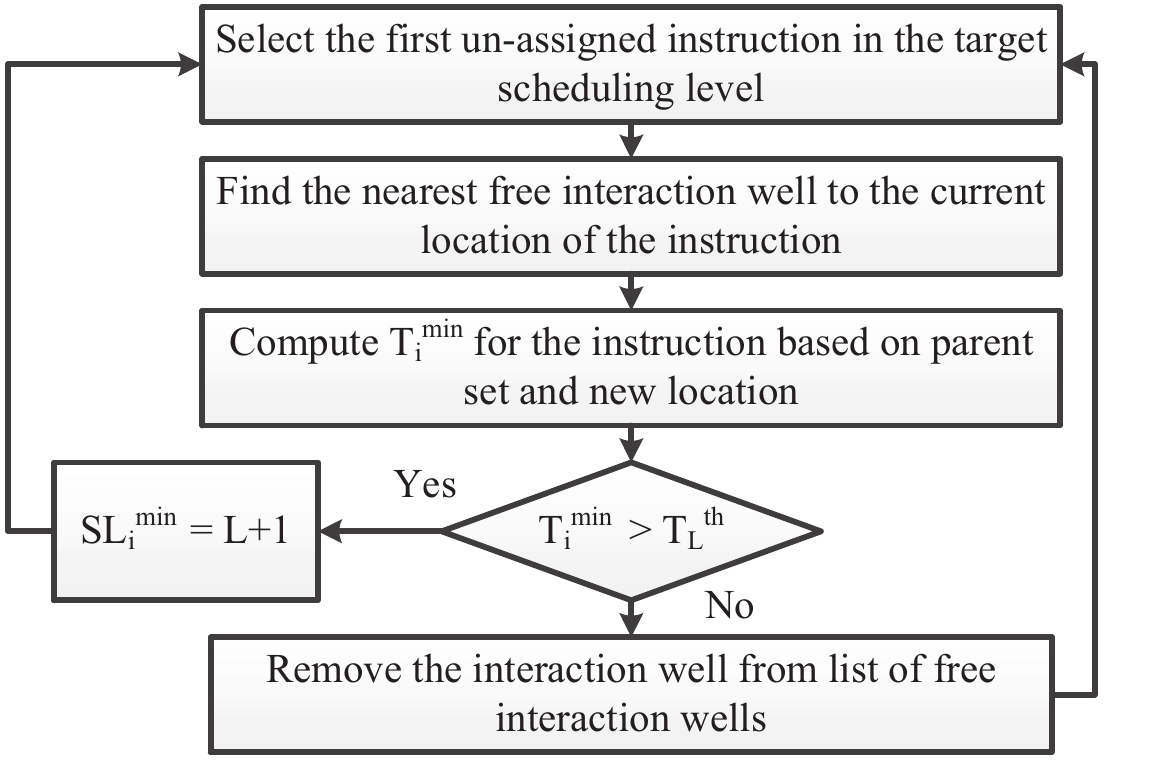}
\caption{Finalization step for instruction placement}
\label{fig:fig11}
\end{figure}

The proposed approach may increase the number of scheduling levels, but it significantly decreases the overall latency of the circuit due to considering a more realistic scheduling solution. The flowchart for finalizing the placement for each scheduling level is provided in fig. \ref{fig:fig11}. 

Notice that repeating the scheduling and global placement after each scheduling level with at least one deferred instruction is costly in case of large circuits. Instead of this approach, the algorithm can be changed to repeat the scheduling and global placement algorithms after the number of deferred instructions reaches a threshold value or the summation of (${T}^{min}_i-T^{th}_{L}$) for deferred instructions reaches a limit. Using this mechanism, the run time of the proposed solution can be shortened at the cost of a slight increase in the resulting circuit latency.

\section{ Experimental Results}
\label{results}

\subsection{ Experimental Setup}

QUFD is implemented in Java. Trapped-ion parameters used for the simulations are as follows: move delay=10$\mu s$; one-qubit primitive instruction delay=50$\mu $s$;$ and two-qubit primitive instruction delay=100$\mu s$. A workstation with Intel Core i7-2600 3.4GHz CPU and 8GB RAM was used to run the simulations.

Determining the concatenation level for the encoding, which is sufficient to get a correct result with high fidelity from the quantum circuit is a process based on the characteristics of the underlying quantum circuit technology and the implemented quantum circuit. This problem is outside the scope of this paper. In this experiment, we focus on one-level Steane code [[7,1,3]] for quantum error correction and the Steane syndrome extraction scheme. Results for some other combinations of encoding and syndrome extraction schemes will also be presented. 

Different ULB sizes are enumerated. For each ULB size, the latency of executing each of the FT operations (CNOT, T, S, X, Y, and Z) encoded with Steane code [[7,1,3]] are found by applying QUFD. Based on the results of the quantum physical mapper and target quantum circuit profiling results, the best ULB size is determined.

\subsection{ Experimental results for different FT operations and different ULB sizes}

The most important part of QUFD is the quantum physical mapper for trapped-ion technology. There are different quantum mapping tools in the literature such as  \cite{whitney_fault_2009}, QUALE \cite{balensiefer_quale:_2005}, QPOS \cite{metodi_scheduling_2006}, and QSPR \cite{dousti_minimizing_2012}. In contrast to our work, all of these approaches rely on ASAP or ALAP scheduling and greedy placement of the qubits and instructions, which in turn results in long circuit latency. QSPR improved on these approaches by repeating the scheduling and placement solution with random initial qubit placement solutions to reach a better initial qubit placement solution, which is critical in case of applying a greedy instruction placement solution. QSPR algorithm provides about 41\% improvement with respect to QUALE. In the following we provide a detailed comparison between QUFD and QSPR \cite{dousti_minimizing_2012}, which is the most advanced mapping tool in the literature.

\begin{figure}
\centering
\includegraphics[width=\textwidth,height=\textheight,keepaspectratio]{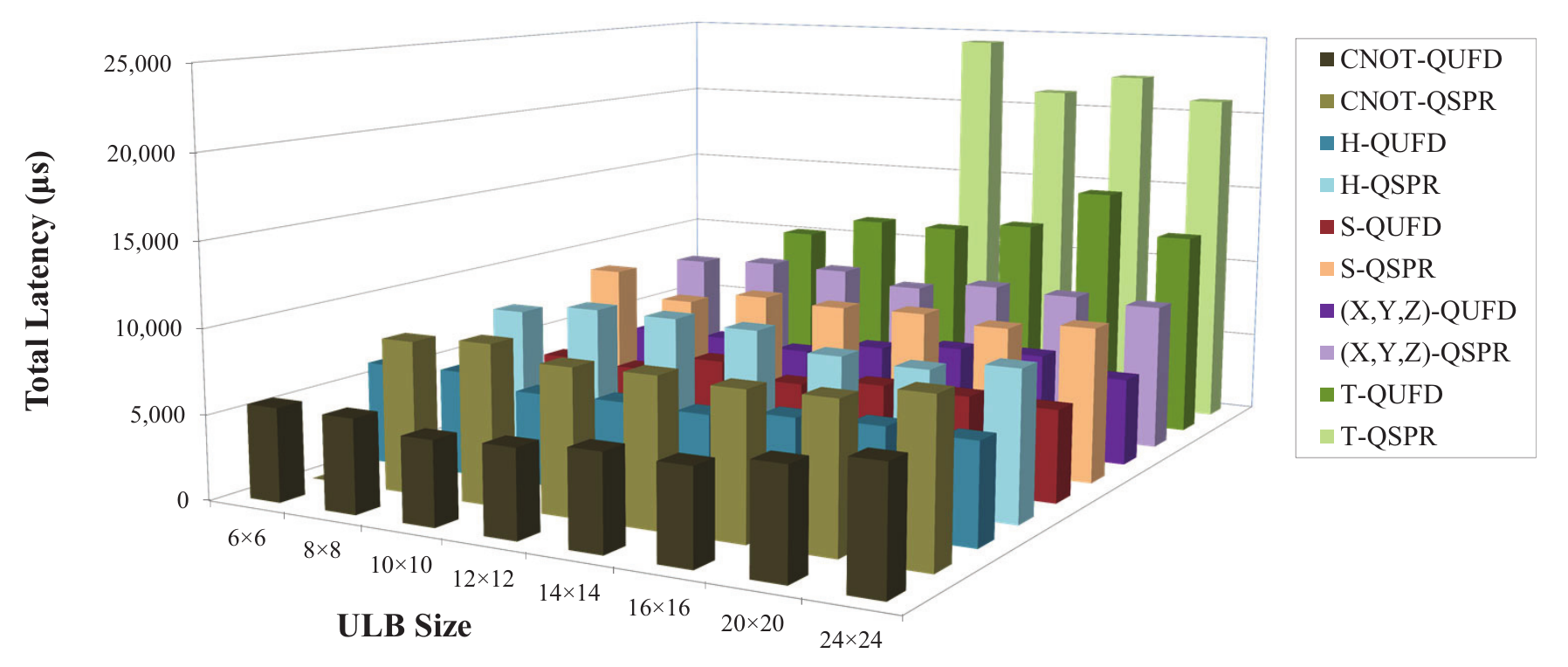}
\caption{Latencies of different FT operations as a function of ULB sizes}
\label{fig:fig12}
\end{figure}

Latencies of various FT operations implemented with different ULB sizes by using QUFD and QSPR are reported in fig. \ref{fig:fig12}. Assuming simultaneous creation of ancillary qubits, there exist no feasible mapping solutions for the T instruction (gate) on 6$\times$6 and 8$\times$8 ULBs. This is because the number of the ancillary qubits for this instruction is more than the number of creation wells in the aforesaid ULBs. QSPR fails to find any mapping solution for the 10$\times$10 and 12$\times$12 ULBs while QUFD succeeds in these cases.

The FT operation latencies obtained by QUFD are from 35\% (for CNOT) to 41\% (for T instruction) lower than those produced by QSPR. Although QUFD uses larger computation time compared to QSPR to produce its solutions, this is not a major concern, because we are only designing ULBs for seven FT operations and this procedure is done only once for the FT operation library design for each quantum circuit fabric.

\begin{figure}
\centering
\includegraphics[width=0.7\textwidth]{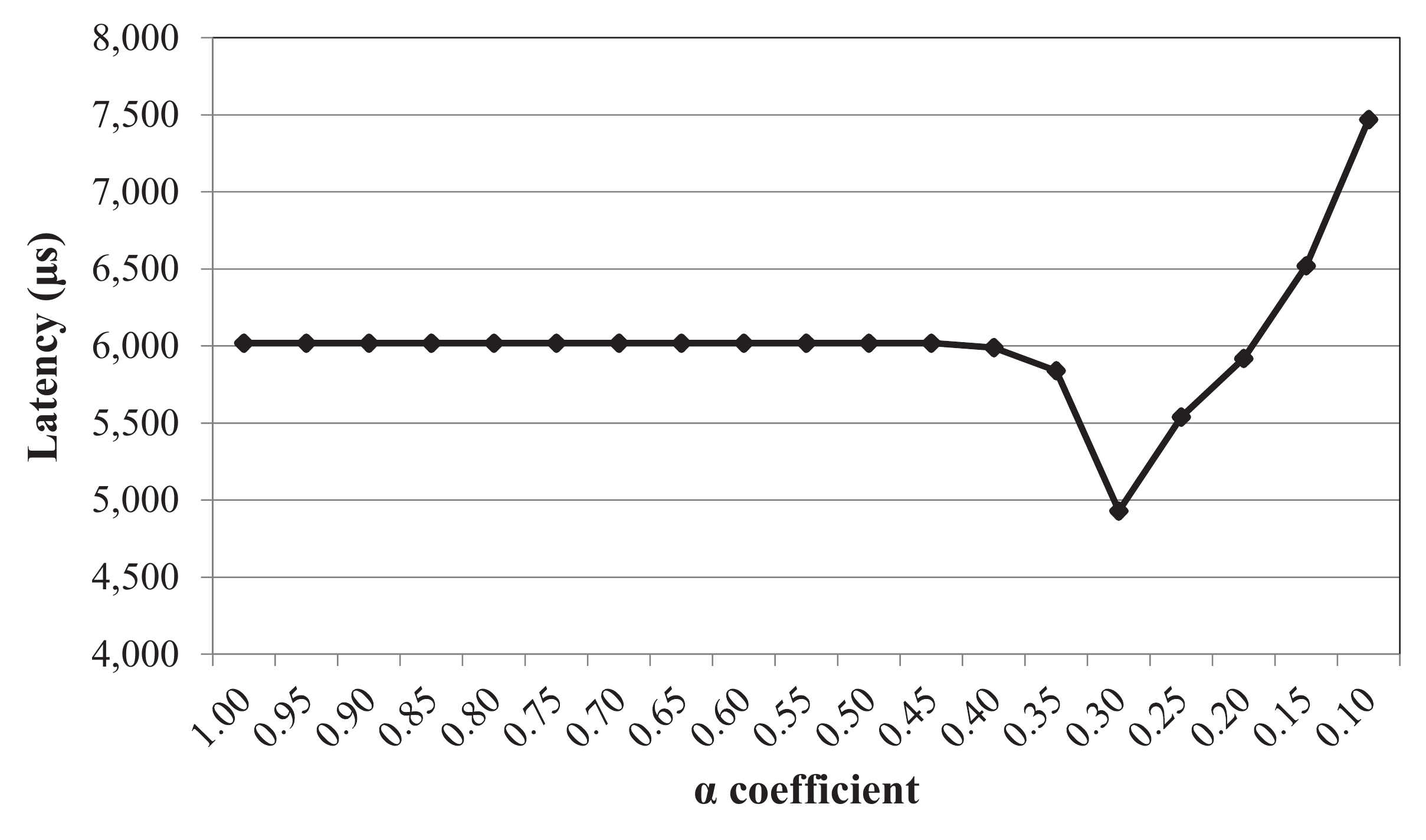}
\caption{Effect of changing ${\mathbf \alpha }$ as the coefficient of $N^{max}$}
\label{fig:fig13}
\end{figure}

As mentioned in section \ref{sched}, tightening the upper-bound on the number of instructions being scheduled in each scheduling level may decrease the total latency of the circuit by decreasing the routing time even though it may increase the number of scheduling levels at the end. This effect is examined for the CNOT instruction mapped to a 10$\times$10 ULB in fig. \ref{fig:fig13}. As can be seen in this figure, if the number of concurrent instructions is limited to 30\% of the physical interaction wells on the fabric, the total latency is decreased by 18\%. It is also seen that decreasing the maximum number of simultaneously active interaction wells does not change the circuit latency until this limit exceeds the maximum number of concurrent instruction in the QIDG ($0.4<\alpha <1$ in fig. \ref{fig:fig13}). Decreasing the maximum number of simultaneously active interaction wells after this value changes the scheduling, placement and routing solution by decreasing the resource contention in each scheduling level for placement and routing problems (positive effect) and possible QIDG depth increase (negative effect). The positive effect is dominant for a part of this process and after that the negative effect starts to dominate. 

\begin{figure}
\centering
\includegraphics[width=0.7\textwidth]{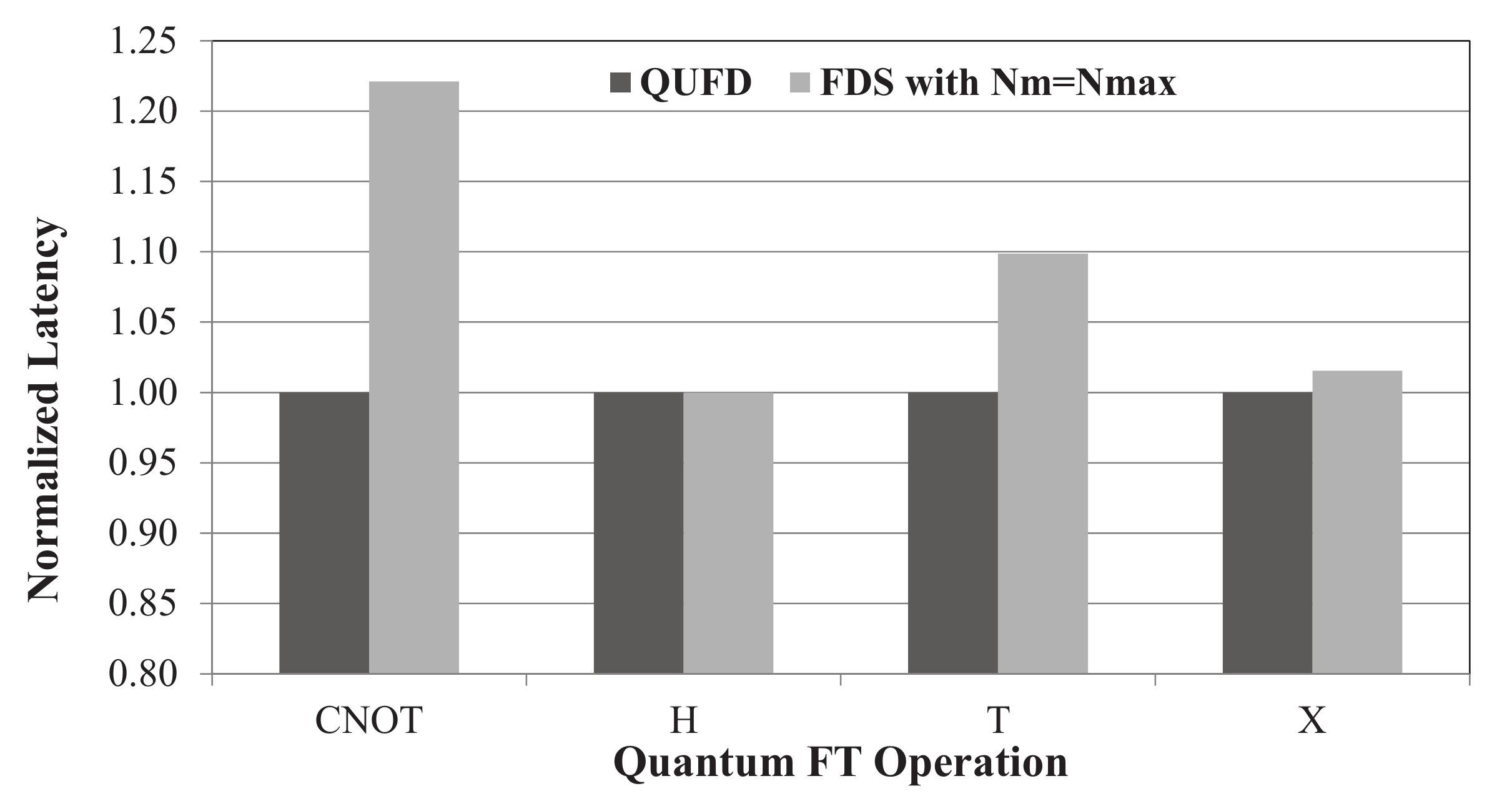}
\caption{Effect of enumeration on ${N^m}$ in the scheduling algorithm on the circuit latency}
\label{fig:fig14}
\end{figure}

To show the effect of enumerating different $N^m$ for the FDS method, (normalized) total latency of the circuit with and without considering this factor for a complete set of FT operations (CNOT, H and T) plus one of the Pauli operations in a 10$\times$10 ULB is reported in fig. \ref{fig:fig14}. As can be seen, for H and X operations, enumeration of different $N^m$ in the scheduling algorithm does not change the total latency but for CNOT and T operations, this enumeration greatly reduces the total latency.   

\begin{figure}
\centering
\includegraphics[width=0.7\textwidth]{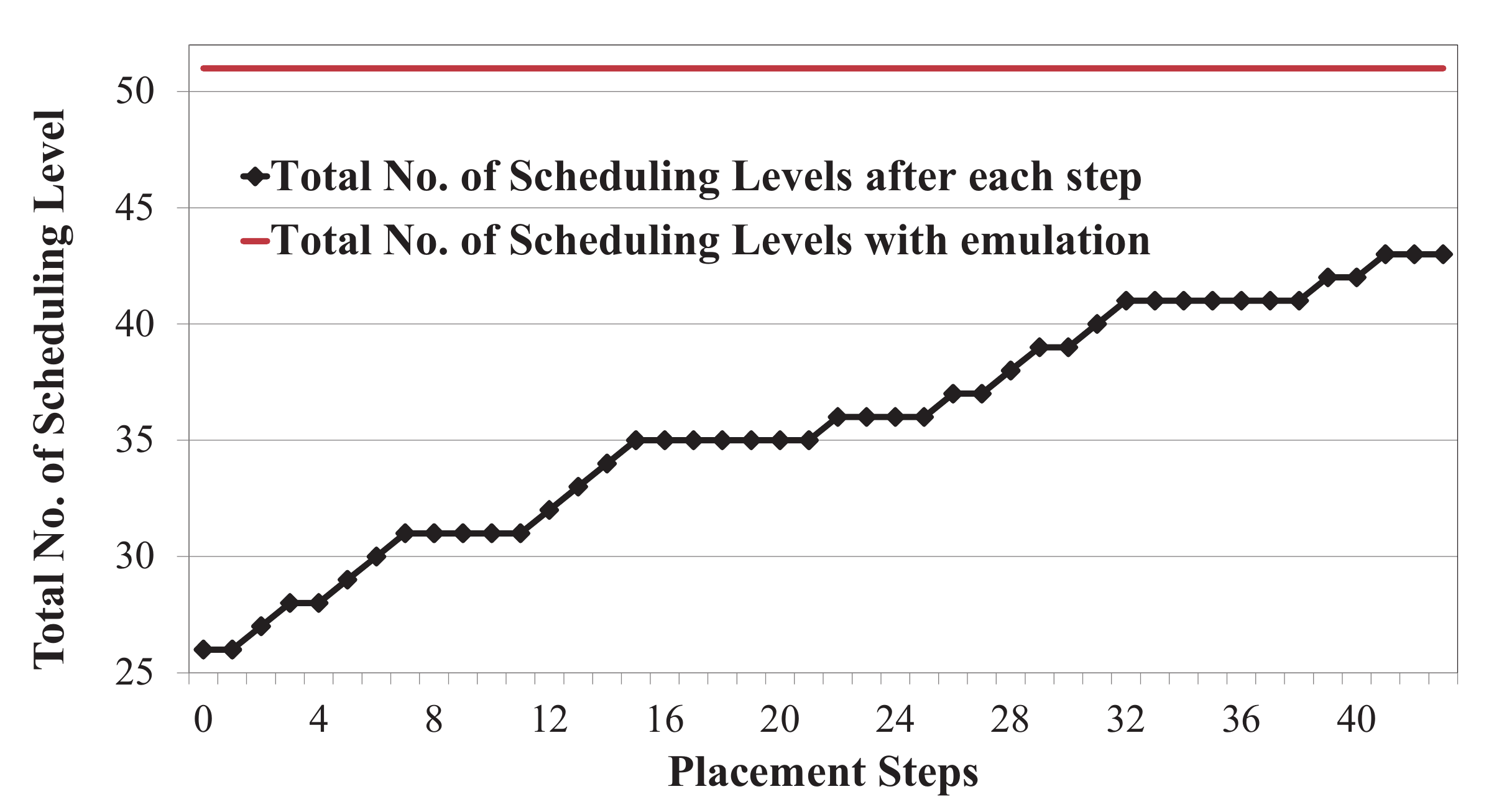}
\caption{Effect of the placement iterations on the number scheduling levels}
\label{fig:fig15}
\end{figure}

Fig. \ref{fig:fig15} shows an exemplary case for the maximum scheduling level in different placement steps for the FT CNOT instruction mapping on a 6$\times$6 ULB. Clearly, as we go further in the placement iterations, the number of scheduling levels is increased when the deferred instruction(s) is on the critical path of the circuit. The actual number of scheduling levels for this case is equal to 51. The predicted number of scheduling levels after scheduling and placement solutions is increased from 26 (without considering the routing latencies) to 43 (after iterative correction of the scheduling levels based on the placement solution). 

\begin{figure}[h]
\centering
\includegraphics[width=0.7\textwidth]{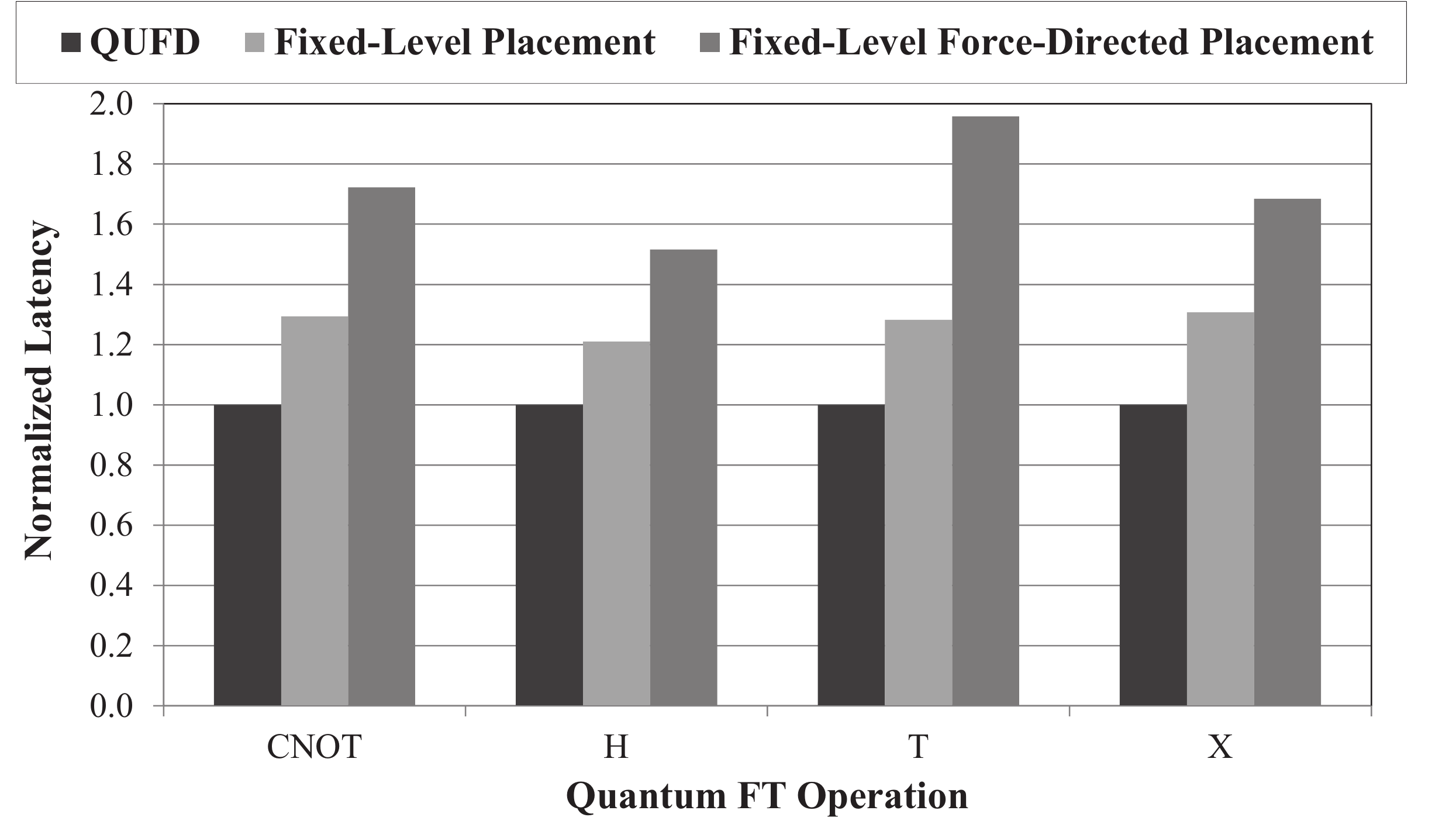}
\caption{Normalized latency of different operations utilizing different placement solutions}
\label{fig:fig16}
\end{figure}

To show the effectiveness of the placement method that utilizes force-directed scheduling to modify the scheduling solution after each placement step, we implemented two other placement methods in the VLSI area. The first algorithm is based on the force-directed placement solution \cite{quinn_forced_1979}. In this approach, each instruction is considered to be a mass and data dependencies in the circuits are modeled by springs. The constant for each spring is set by the criticality factor of the source and the destination qubits. To avoid overlap between instructions with the same scheduling level, repulsive forces are used. The second algorithm is based on SimPL algorithm \cite{kim_simpl:_2010} with fixed scheduling levels for instructions. Normalized latencies (w.r.t. QUFD results) for a complete set of FT operations (CNOT, H and T) plus one of the Pauli operations in a 12$\times$12 ULB for different placement mechanisms are presented in fig. \ref{fig:fig16}.

As can be seen, the proposed placement method that modifies the scheduling level of instructions based on the placement solution generates on average 27\% better results than the same placement method without using adaptive scheduling levels. Moreover, the placement method based on the force-directed placement generates on average 72\% worse results compared to the proposed placement solution.

\begin{figure}[h]
\centering
\includegraphics[width=0.7\textwidth]{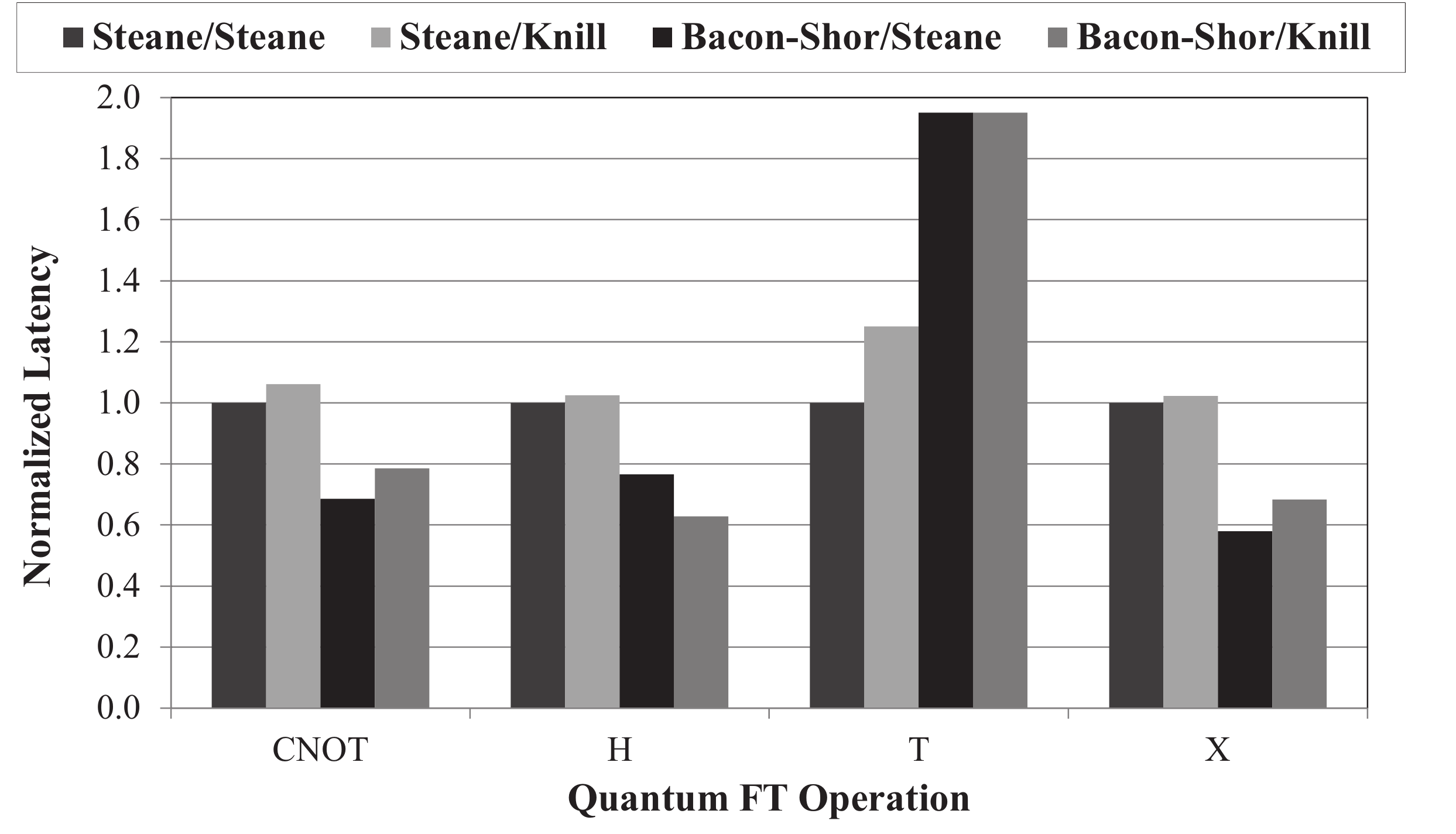}
\caption{Normalized latency of different operations in different encoding/syndrome extraction schemes}
\label{fig:fig17}
\end{figure}

In this experiment, we focus on one-level Steane encoding [[7,1,3]] and Steane syndrome extraction scheme but to show the difference between encoding and syndrome extraction schemes, results of another concatenated code, that is, Bacon-Shor with the lowest concatenation level and different syndrome extraction schemes (Steane and Knill ancilla) are reported in fig. \ref{fig:fig17}.

In this figure, latencies of CNOT, H and X operations are measured in a 12$\times$12 ULB. For the T operation, a 20$\times$20 ULB is used because the number of required ancillary qubits for the T gate for some of the schemes is higher than the number of available creation wells on a 12$\times$12 ULB. Note, however, that these encoding/syndrome extraction schemes provide different levels of error detection/correction capabilities for the quantum circuit and cannot be compared solely based on these latencies.

\subsection{ Selecting the best ULB size}

\begin{figure}[h]
\centering
\includegraphics[width=0.7\textwidth]{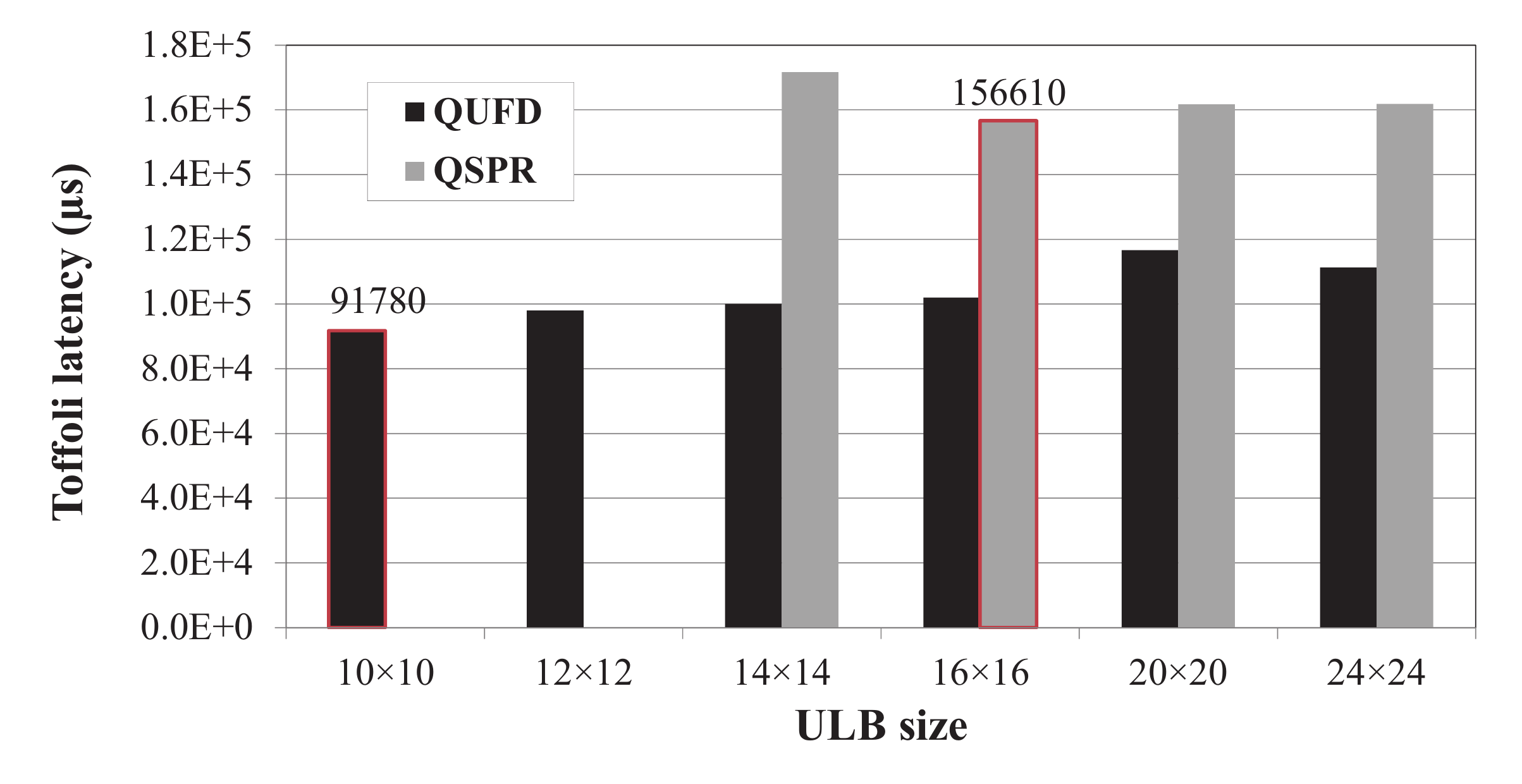}
\caption{Latency of Toffoli circuit for different ULB sizes}
\label{fig:fig18}
\end{figure}

Based on the found latencies for FT operations, the total latency of the mapped Toffoli circuit to a 2$\times$2 mesh TQA, which is calculated in section \ref{cadflow}, is reported in fig. \ref{fig:fig18}. Note that results for the 10$\times$10 and 12$\times$12 ULB sizes are not reported for QSPR because this tool fails to find the total latency of the T instruction for these ULB sizes.

As can be seen, the best ULB size for QUFD is 10$\times$10 and the best ULB size for QSPR is 16$\times$16. Moreover, the best total latency for the proposed QUFD is 41\% less than the best total latency of QSPR.

\section{ Conclusion}
\label{concl}

This paper presented quantum scheduling and placement algorithms for mapping fault-tolerant quantum operations to a universal logic block (ULB). Analytical formulations for the scheduling and placement problems were also presented. Effective and state-of-the-art algorithms used in standard CAD tools for classical circuits were modified to be applicable to the quantum mapping problem (a quantum CAD problem). In particular, modified list scheduling and force-directed scheduling were applied to the quantum instruction scheduling problem to capture the no-cloning dependencies among quantum instructions and determine the scheduling levels to minimize the number of instruction levels and reduce the resource contention. To consider the large impact of the qubit routing on the total latency of the quantum circuit, a net-weighting timing-driven placement solution based on SimPL package was employed. This technique adaptively modifies the scheduling level of the instructions. Moreover, a technique was described to determine the best ULB size for implementing FT quantum operations. All of the aforementioned algorithms were implemented in a CAD tool called QUFD. Experimental results showed that QUFD dominated the prior art quantum mappers by as much as 41\% in terms of the average latency reduction. 

With some modifications, QUFD can be applied to large QODG's to produce a mapping solution for any tile-based quantum architectures. Moreover, the proposed flow is applicable to quantum mapping and ULB design in other quantum fabric technologies with small changes due to different qubit/information movement cost. 

{\bf Acknowledgment}: This research was supported by the Intelligence Advanced Research Projects Activity (IARPA) via Department of Interior National Business Center contract number D11PC20165. The U.S. Government is authorized to reproduce and distribute reprints for Governmental purposes notwithstanding any copyright annotation thereon. The views and conclusions contained herein are those of the authors and should not be interpreted as necessarily representing the official policies or endorsements, either expressed or implied, of IARPA, DoI/NBC, or the U.S. Government.



%
%

\end{document}